\documentclass[pdflatex,sn-mathphys-num]{sn-jnl}

\usepackage{graphicx}%
\usepackage{multirow}%
\usepackage{amsmath,amssymb,amsfonts}%
\usepackage{amsthm}%
\usepackage{mathrsfs}%
\usepackage[title]{appendix}%
\usepackage{xcolor}%
\usepackage{textcomp}%
\usepackage{manyfoot}%
\usepackage{booktabs}%
\usepackage{float}
\usepackage{algorithm}%
\usepackage{algorithmicx}%
\usepackage{algpseudocode}%
\usepackage{listings}%
\usepackage{threeparttable}
\usepackage{verbatim}
\usepackage{tabularx}
\usepackage{array}
\usepackage[normalem]{ulem}

\newcolumntype{Y}{>{\raggedright\arraybackslash}X}

\theoremstyle{thmstyleone}%
\theoremstyle{thmstyletwo}%

\theoremstyle{thmstylethree}%

\begin{document}

\title[Article Title]{Real-to-Sim Calibration of Bosch Etching Simulators with Few-Shot XSEM Data and Hybrid Search}


\author*[1]{\fnm{Xinyu} \sur{Jia}}\email{jia181@purdue.edu}

\author[1,2]{\fnm{Jainil} \sur{Shah}}\email{shah1206@purdue.edu}

\author[3]{\fnm{Rahul} \sur{Tripathi}}\email{itsrahultripathi@gmail.com}

\author[1]{\fnm{Vijay} \sur{Gupta}}\email{gupta869@purdue.edu}

\author[1,2]{\fnm{Zhihong} \sur{Chen}}\email{zhchen@purdue.edu}

\affil*[1]{%
  \orgdiv{Elmore Family School of Electrical and Computer Engineering},
  \orgname{Purdue University},
  \orgaddress{%
    \city{West Lafayette},
    \state{Indiana},
    \country{USA}
  }
}

\affil[2]{%
  \orgname{Birck Nanotechnology Center},
  \orgname{Purdue University},
  \orgaddress{%
    \city{West Lafayette},
    \state{Indiana},
    \country{USA}
  }
}

\affil[3]{%
  \orgname{Micron Semiconductor Asia Operations Pte. Ltd.},
  \orgaddress{%
    \street{1 North Coast Drive},
    \city{Singapore},
    \postcode{757432},
    \country{Singapore}
  }
}


\abstract{The Bosch process is widely used in MEMS and TSV fabrication, but accurate prediction of etching outcomes remains difficult without calibration against real measurements, which are sparse and costly to obtain. This work presents a few-shot real-to-sim calibration framework that decouples simulator parameters from direct process variables to infer a recipe-specific simulator parameter vector from sparse measurements, enabling reliable prediction of etch depth and sidewall angle under held-out process conditions. We first formulate Bosch profile prediction as a few-shot real-to-sim calibration problem, where a small set of real XSEM measurements is used to identify a recipe-specific ViennaPS parameter vector $z_r^*$ that minimizes the discrepancy between simulated and measured etch depth and sidewall angle. To search this parameter space efficiently, the framework combines local exploration, LLM-guided candidate generation, and XGBoost-based surrogate screening, while retaining ViennaPS simulation as the final verification step for each selected candidate. Experiments on two Bosch etching recipes show that the proposed framework accurately predicts held-out etching results, achieving normalized global errors of 0.340 and 0.249 for Recipe 1 and Recipe 2 respectively.}

\keywords{Bosch process, Real-to-simulation, Few-shot learning, Physics-based process simulation}



\maketitle

\section{Introduction}\label{introduction}

Deep Reactive Ion Etching (DRIE) is a critical process for Microelectromechanical Systems (MEMS) and Through-Silicon Vias (TSV) in 3D packaging. Among DRIE techniques, Bosch process is widely used in industrial manufacturing because it can produce deep silicon etching profiles. Since etching outcomes directly affect device performance and process yield, accurate prediction is essential for process design and optimization. However, physical Bosch etching experiments are expensive and time-consuming, which makes predictive process models necessary for estimating etching outcomes before running new experiments. Prediction alone, however, is not sufficient. A model may appear accurate in simulation, but if it is not calibrated against fab measurements, its practical value is limited. The key challenge is therefore calibration, which closes the gap between model outputs and experimental behavior when real data are sparse and expensive to collect.

Several modeling strategies have been explored for etching prediction. By reconstructing the process in a virtual environment, physics-based process simulators (e.g., ViennaPS\cite{reiter2025viennaps}, Synopsys\cite{goldman2009synopsys}, Silvaco\cite{sarkar2018device}) can predict etch profiles without consuming expensive wafers. However, their predictive accuracy depends on internal simulator parameters that are not directly determined by equipment settings. By directly fitting experimental measurements, machine learning (ML) models can learn complex relationships between process inputs and outputs without relying on an explicit physical model, but its success usually depends on sufficient labeled data, which is difficult for cross-sectional scanning electron microscopy(XSEM)-based Bosch etching studies. Physics-Informed Neural Networks (PINNs) reduce data dependence by incorporating known physical equations into the learning objective. For Bosch etching, physics-based equations can represent the dominant transport and surface-reaction mechanisms, but their effective parameters are difficult to specify accurately in advance and often vary from one tool or recipe to another. Therefore, models based on fixed physical constraints may not fully reproduce measured etch profiles without calibration against real data.

These limitations motivate a real-to-sim calibration problem in which simulator parameters are decoupled from direct process variables. Real-to-simulation calibration has been extensively studied in robotics, where limited real measurements can be used to estimate simulator parameters that align virtual and physical system behavior. However, most of these methods rely on continuous information such as image sequences, object trajectories, or interaction rollouts. Our setting provides only sparse XSEM measurements after the etching process completes, namely etch depth and sidewall angle. Therefore, existing robotics-oriented real-to-sim methods are not directly designed for our setting,where simulator calibration must be performed using only sparse terminal XSEM measurements rather than continuous interaction observations.

To address this gap, we present a few-shot real-to-sim simulator calibration framework for reliable Bosch etching prediction. Specifically, a small number of real experimental measurements are used to guide the identification of ViennaPS simulator parameters, aligning the simulated etch depth and sidewall angle with real Bosch etching results. After calibration, the simulator is no longer used only for simulation. It becomes a physics prior that is aligned with real experimental behavior. This enables more reliable prediction of etching outcomes under held-out process conditions within the same recipe. The main contributions of this work are summarized as follows:

\begin{itemize}
    \item We formulate Bosch etching prediction as a few-shot simulator-parameter identification problem. Given a small set of real XSEM measurements for each recipe \(r\), the proposed framework identifies a recipe-specific ViennaPS parameter vector $z_r^*$ by decoupling simulator parameters from direct process variables and jointly optimizing $z_r^*$ across multiple fitting conditions under the same recipe. After calibration, ViennaPS remains an executable physics-based simulator, but its predicted etch depth and sidewall angle are better aligned with real Bosch etching behavior. This enables more reliable prediction for held-out process conditions without retraining a black-box model or requiring test samples during calibration.
    \item We develop a hybrid search strategy for efficient simulator-parameter calibration. In the early stage, when few simulation records are available, the LLM module relies on Bosch process knowledge and current residuals to propose physically plausible candidates. As the calibration history grows, an XGBoost surrogate is trained to approximate the ViennaPS forward response and pre-screen a larger candidate pool, reducing the number of expensive full simulations needed. The surrogate serves only as a filter and the final parameter update is always determined by ViennaPS evaluation on $D_{\mathrm{fit}}$. This design combines physics guided cold start exploration, data driven screening, and simulation based verification to reduce calibration cost.
\end{itemize}

The remainder of this paper is organized as follows. Section~\ref{related_work} reviews related work in etching prediction and simulation, and real-to-simulation calibration. Section~\ref{Prob_formul} shows the Bosch etching process and presents the problem formulation, where simulator calibration is defined as identifying a physics-based simulator parameter with few-shot real XSEM measurements. Section~\ref{sec:Method} introduces the overall real-to-sim framework. Section~\ref{sec:Experiments} reports the experimental results and evaluates the calibrated simulator in terms of fitting quality and prediction ability. Finally, Section~\ref{sec:conclusion} concludes the paper and discusses future work.

\section{Related Work}\label{related_work}

\subsection{Etching Prediction and Simulation}
\label{sec:etching_prediction}

In this section, we discuss existing approaches for etching prediction and simulation, including physics-based simulators, ML models, and PINNs. As summarized in Table~\ref{exist_difference}, these approaches respectively require calibration of unknown simulator parameters, large amounts of real data, or sufficiently complete governing equations.
\begin{table*}[t]
\centering
\caption{Comparison of existing approaches for etching prediction and simulation.}
\label{exist_difference}

\footnotesize
\setlength{\tabcolsep}{5pt}
\renewcommand{\arraystretch}{1.15}

\begin{tabularx}{\textwidth}{
    @{}
    p{2.5cm}
    X
    X
    @{}
}
\toprule
\textbf{Approach}
&
\textbf{Modeling basis and data requirement}
&
\textbf{Limitation for Bosch etching}
\\
\midrule

Physics-based simulator
\cite{reiter2025viennaps,bobinac2023effect,rodrigues20233d,
synopsys_sentaurus_topo,silvaco_victory}
&
Uses explicit physical models of particle transport, surface reactions, and profile evolution.Calibration against measured data is required to determine internal simulator parameters.
&
The internal parameters cannot be directly obtained from experimental recipes and may vary across process conditions.
\\

ML models
\cite{kang2026machine,kim2025adaptive,breidung2025process}
&
Learn mappings from process inputs or sensor measurements to process outcomes, typically using large labeled datasets or extensive sensor data.
&
Real XSEM measurements are scarce, while purely data-driven models provide limited physical interpretability under small-data settings.
\\

PINNs
\cite{dighamber2024physics,zhong2022low,leng2025physics}
&
Combine training data with governing equations by embedding physical constraints into the learning objective.
&
Existing methods rely on prescribed governing equations or simulation data, but do not calibrate simulator parameters against real measurements.
\\

\bottomrule
\end{tabularx}
\end{table*}

\paragraph{Physics-based simulators}
Physics-based process simulators have been widely adopted for modeling semiconductor fabrication. ViennaPS\cite{reiter2025viennaps} is a physics-based process simulator providing a flexible framework for feature-scale topography evolution in semiconductor fabrication. Its level-set surface representation and Monte Carlo ray-tracing capability enable simulation of profile evolution driven by particle fluxes and surface reactions. Bobinac et al.\cite{bobinac2023effect} implement an $\text{SF}_6/\text{O}_2$ plasma etching model within ViennaPS to study the influence of mask geometry on etch profiles. Rodrigues et al.\cite{rodrigues20233d} model the three-dimensional feature-scale etching of $\text{SiO}_2$ under $\text{CF}_4/\text{C}_4\text{F}_8$ fluorocarbon plasma. Commercial tools such as Synopsys Sentaurus Topography\cite{synopsys_sentaurus_topo} and Silvaco Victory Process\cite{silvaco_victory} further support profile evolution simulation. Therefore, physics-based simulators can supply valuable physical priors for etch profile prediction. However, their predictive accuracy depends critically on internal model parameters that cannot be read directly from experimental recipes and vary with process conditions. Consequently, although physics-based simulators reduce experimental cost and yield physically interpretable predictions, real-to-sim calibration against measured data remains necessary to match a target Bosch etching condition.
\paragraph{ML models}
Data-driven ML models are a common approach for process modeling and outcome prediction in semiconductor manufacturing. Kang et al.\cite{kang2026machine} showed that neural network models can predict etch depth from process parameters or RGB colorimetric features, demonstrating the potential of data-driven model for plasma etching.
Kim et al.\cite{kim2025adaptive} address chamber-to-chamber variation in plasma etching by applying adversarial domain adaptation to achieve robust endpoint detection without labeled data from target chambers.
Breidung et al.\cite{breidung2025process} train ML models on process sensor data from a high-volume CMP fab to predict wafer-level material removal rates and surface non-uniformity, establishing a virtual metrology pipeline without additional metrology overhead. However, these data-driven methods require large number of labeled real samples and learn a black-box mapping from process inputs to outputs without capturing the physical structure of etch-profile evolution. This makes them less suitable for Bosch etching prediction, where real XSEM measurements are costly and experimental samples are scarce.
\paragraph{PINNs}

By embedding physical constraints into the learning objective, PINNs can improve data efficiency and provide more physically consistent predictions. Dighamber~\cite{dighamber2024physics} trained a PINN using plasma etch simulations generated by ViennaPS and incorporated a physics loss based on the level-set equation to improve the physical consistency of predicted etch-profile evolution. However, this method learns from simulated data and does not calibrate ViennaPS parameters against real measurements. Zhong et al.\cite{zhong2022low} propose CS-PINN and RK-PINN to solve plasma governing equations by embedding physical constraints directly into the neural network loss, demonstrating that PINNs can effectively handle complex non-linear plasma dynamics with limited data.
Leng et al.\cite{leng2025physics} provide a comprehensive review of physics-informed machine learning and PINNs in intelligent manufacturing, showing that embedding physical constraints into ML training improves data efficiency, accuracy, and generalization in settings where large datasets are difficult to obtain. However, existing PINN approaches either rely on prescribed governing equations or learn surrogate models from simulation data. Physics-based models can capture the main transport and surface-reaction mechanisms in Bosch etching, but the full plasma chemistry and its effective parameters are hard to specify precisely for a given tool and recipe. As a result, fixed physics constraints alone cannot fully capture the measured process behavior. This motivates calibrating an approximate physics-based simulator using sparse real measurements, rather than assuming that the governing equations or the simulator parameters are known exactly.

\subsection{Real-to-Sim Calibration}
\label{sec:real2sim}
Recent sim-to-real studies mainly aim to improve prediction on real data by transferring knowledge from simulations. Rautela et al. \cite{rautela2026sim+} jointly train on simulation and experimental data to maintain balanced performance across both domains. In contrast, our goal is to use sparse real measurements to calibrate the parameters of a physics-based simulator. Thus, our problem is better characterized as real-to-sim calibration.

Real-to-sim calibration aims to use real-world observations to adjust simulator parameters so that the simulator can better reproduce real system behavior. Existing real-to-sim studies have been largely developed in robotics, where limited real world interaction data are used to estimate simulator parameters and improve simulation fidelity.
Chebotar et al.\cite{chebotar2019closing} adapt simulation randomization parameters using a small number of real-world robot rollouts, demonstrating that limited real-world interactions are sufficient to close the sim-to-real gap in robotic manipulation tasks.  
Torne et al.\cite{torne2024reconciling} propose RialTo, a real-to-sim-to-real pipeline that constructs digital twin simulation environments from small amounts of real-world data and uses reinforcement learning to fine-tune imitation learning policies.
Baumeister et al.\cite{baumeister2025incremental} propose a real-to-sim approach for robotic object pushing that incrementally calibrates physics simulator parameters from a few real-world interaction rollouts using sampling-based optimization, enabling improved model-predictive control with minimal real-world data.

In manufacturing, real-to-sim approaches have also been explored for system-level simulation.
Behrendt et al.\cite{behrendt2026real} apply a real-to-sim approach in semiconductor manufacturing, using production data to automatically generate simulation models for a digital twin without manual modeling effort.

These studies demonstrate that limited real-world data can be effective for improving simulation fidelity. However, all of them rely on continuous observations. This assumption does not hold in our Bosch etching setting, where no intermediate process states are observable during etching. Instead, each real sample only provides sparse terminal XSEM measurements after the process is completed. As a result, their optimization strategies cannot be directly applied here. We therefore need a real-to-sim calibration method designed specifically for feature-level Bosch etching simulation.

\section{Background and Problem Formulation}
\label{Prob_formul}

\subsection{Bosch Etching Process}
\label{bosch_process}

The Bosch process is a cyclic deep reactive ion etching (DRIE) technique used to fabricate high-aspect-ratio silicon structures. It alternates between fluorine-based silicon etching and fluorocarbon-based sidewall passivation. In the process considered in this work, an initial \(\mathrm{SF}_6/\mathrm{Ar}\) ignition stage exposes the silicon surface and initiates trench formation. Each subsequent cycle begins with a \(\mathrm{C}_4\mathrm{F}_8\)-based deposition step, during which a polymer passivation layer is formed on the trench bottom and sidewalls. This is followed by two \(\mathrm{SF}_6\)-based etching stages. Etch A preferentially removes the passivation layer from the trench bottom through directional ion bombardment, while Etch B etches the exposed silicon and extends the trench vertically. Repeating this deposition, passivation-removal, and silicon-etching sequence protects the sidewalls while enabling deep anisotropic etching, resulting in the characteristic scalloped profile of the Bosch process.


\subsection{Real-to-Sim Calibration Formulation}
\label{calibration_formulation}

In this section, we formulate real-to-sim simulator calibration as identifying a shared ViennaPS parameter vector that aligns simulated etching outcomes with real Bosch etching behavior. 
After calibration, the simulator can be used as a physics based predictive model for estimating etch depth and sidewall angle under new process conditions in the calibrated recipe setting.

For the $r$-th recipe, the Bosch process input is defined by $x^{(r)} \in \mathcal{X}$, where $x^{(r)} = \left(c, o, p_r\right)$. Here, $c$ is the number of cycles in the Bosch process, $o$ is the opening size, and $p_r \in \mathbb{R}^k$ is the process parameter vector of the $r$-th recipe.  

Let $z \in \mathcal{Z}$ denote the simulator parameter vector to be identified. Given a process input $x$ and a simulator parameter vector $z$, ViennaPS produces a simulation output $s = S(x; z)$, where $S(\cdot; \cdot)$ represents the simulator and $s$ denotes the simulated profile.  To quantitatively evaluate the profile, we extract etch depth and sidewall angle as characterization metrics:
\begin{equation}
    \phi(s) = \begin{bmatrix} d_{sim} \\ \alpha_{sim} \end{bmatrix} \in \mathbb{R}^2,
\end{equation}
where $d_{sim}$ is the simulated etch depth and $\alpha_{sim}$ is the simulated sidewall angle.

Correspondingly, the real etching profile \(y = Y(x)\) is similarly characterized by:
\begin{equation}
    \phi(y)= \begin{bmatrix} d_{real} \\ \alpha_{real} \end{bmatrix} \in \mathbb{R}^2,
\end{equation}
where $d_{real}$ and $\alpha_{real}$ represent the measured etch depth and sidewall angle, respectively.


Due to uncertain simulator parameters and simplified physical assumptions, the raw simulation output may not directly match real experimental measurements. Therefore, our goal is to identify a simulator parameter vector $z$ that aligns ViennaPS simulation with real Bosch etching behavior.

Given the few-shot real experimental dataset
\begin{equation}
\mathcal{D}^{(r)}_{\mathrm{real}}=\{(x^{(r)}_i,y^{(r)}_i)\}_{i=1}^{N^{(r)}},
\end{equation}
where $i$ indexes the real experimental samples within recipe $r$.

For each recipe \(r\), we divided the real observations into two non-overlapping sets:
\begin{equation}
\label{eq:realset}
\mathcal{D}_{\mathrm{real}}^{(r)}
=
\mathcal{D}_{\mathrm{fit}}^{(r)}
\cup
\mathcal{D}_{\mathrm{test}}^{(r)},
\qquad
\mathcal{D}_{\mathrm{fit}}^{(r)}
\cap
\mathcal{D}_{\mathrm{test}}^{(r)}
=
\emptyset .
\end{equation}

Here, \(\mathcal{D}_{\mathrm{fit}}^{(r)}\) is used to identify the shared simulator parameter vector \(z_r^*\), while \(\mathcal{D}_{\mathrm{test}}^{(r)}\)  is used only for final evaluation. We define
\[
N_{\mathrm{fit}}^{(r)}
=
|\mathcal{D}_{\mathrm{fit}}^{(r)}|,
\qquad
N_{\mathrm{test}}^{(r)}
=
|\mathcal{D}_{\mathrm{test}}^{(r)}|,
\qquad
N_{\mathrm{real}}^{(r)}
=
|\mathcal{D}_{\mathrm{real}}^{(r)}|.
\]

Our goal is to identify a recipe-specific simulator parameter vector $z_r^*$ that aligns ViennaPS simulation with real Bosch etching behavior across the considered process conditions within recipe r. This can be formulated as
\[
z_r^*
=
\arg\min_{z\in\mathcal{Z}}
\frac{1}{N^{(r)}_{\mathrm{fit}}}
\sum_{(x_i^{(r)},y_i^{(r)})\in \mathcal{D}^{(r)}_{\mathrm{fit}}}
\mathcal{L}\left(\phi(S(x_i^{(r)};z)),\phi(y_i^{(r)})\right).
\]
Here a separate parameter vector \(z_r^*\) is identified for each recipe, while the same \(z_r^*\) is shared across all fitting conditions within that recipe.


After calibration,  for a test condition \( (x_j^{(r)}, y_j^{(r)}) \in \mathcal{D}_{\mathrm{test}}^{(r)}\), the calibrated simulator for recipe \(r\) first generates a predicted simulated profile:
\begin{equation}
\hat{s}_j^{(r)}
=
S\left(x_j^{(r)}; z_r^*\right).
\end{equation}

The predicted etch measurements are then obtained by applying the profile measurement operator:

\begin{equation}
\hat{m}_j^{(r)}
=
\phi\left(\hat{s}_j^{(r)}\right)
=
\begin{bmatrix}
\hat{d}_j^{(r)} \\
\hat{\alpha}_j^{(r)}
\end{bmatrix}.
\end{equation}

These predicted measurements are compared with the corresponding real measurements

\begin{equation}
m_j^{\mathrm{real},(r)}
=
\phi\left(y_j^{(r)}\right)
=
\begin{bmatrix}
d_j^{\mathrm{real},(r)} \\
\alpha_j^{\mathrm{real},(r)}
\end{bmatrix}.
\end{equation}

For a new process condition $x_{\mathrm{new}}$, the predicted etching outcome is
\begin{equation}
\hat{s}^{(r)}_{\mathrm{new}}= S(x_{\mathrm{new}^{(r)}};z_r^*),
\end{equation}

The corresponding predicted etch depth and sidewall angle are
\begin{equation}
\hat{m}_{\mathrm{new}}^{(r)}
=
\phi(\hat{s}_{\mathrm{new}}^{(r)})
=
\begin{bmatrix}
\hat{d}^{(r)}_{\mathrm{new}} \\
\hat{\alpha}^{(r)}_{\mathrm{new}}
\end{bmatrix}.
\end{equation}

In summary, for each recipe \(r\), few-shot real experimental data are used to calibrate a recipe-specific ViennaPS parameter vector $z_r^*$. Then, the calibrated simulator serves as the real-to-sim bridge for predicting etch depth and sidewall angle under held-out process conditions within the same recipe setting.

\section{Methods}
\label{sec:Method}
\subsection{Real-to-Sim Calibration Framework}

\begin{figure}[t]
    \centering
    \includegraphics[width=1.0\textwidth]{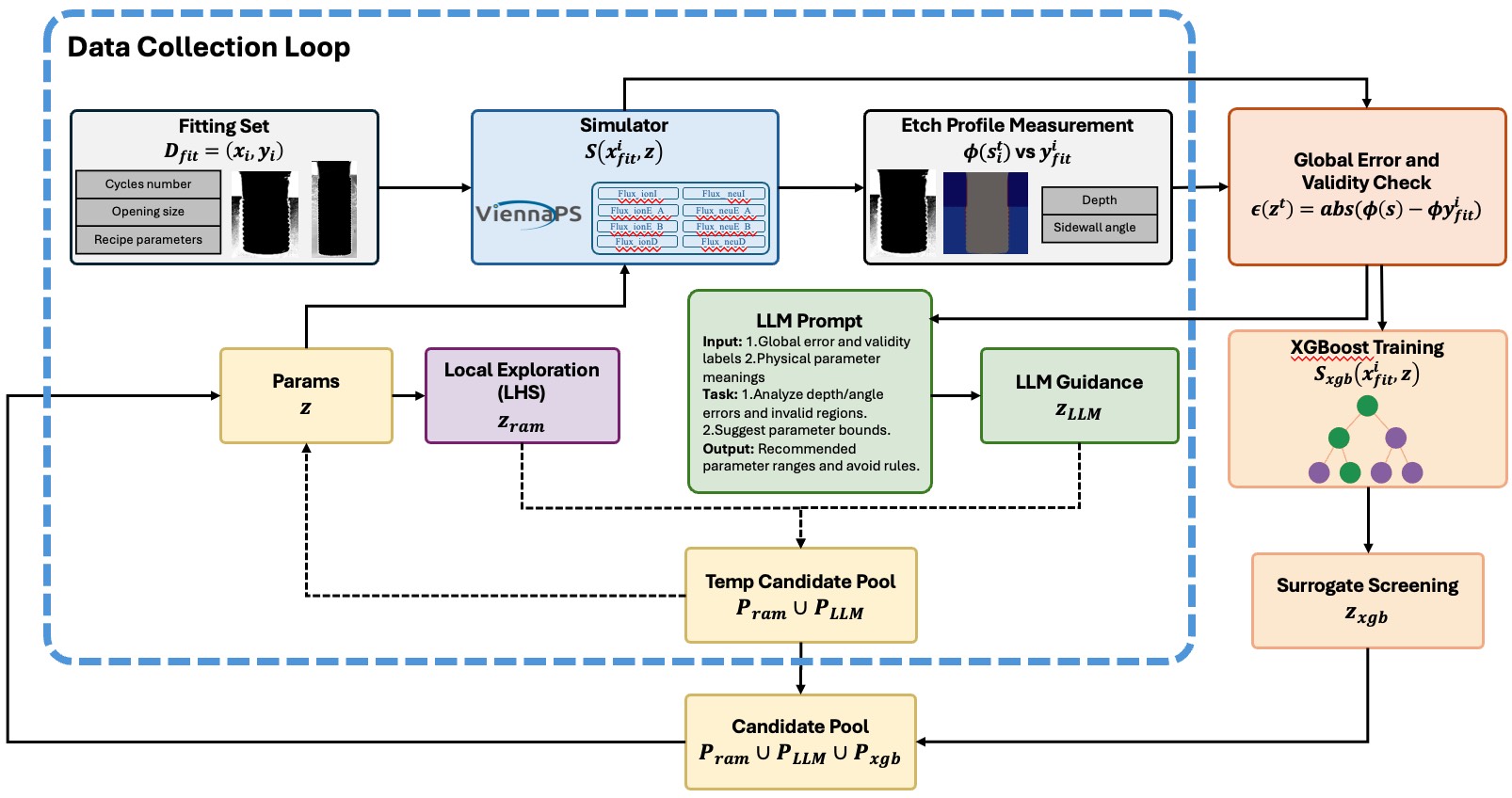}
    \caption{Parameter Identification Framework}
    \label{fig:framework}
\end{figure}

Fig.~\ref{fig:framework} shows the overall framework of the proposed calibration method. For each recipe \(r\), the goal is to identify a shared ViennaPS simulator parameter vector $z_r^*$ using a few-shot real experimental measurements. After calibration, the simulator is used to generate predicted profiles for held-out process conditions, and the same measurement operator \(\phi(\cdot)\) is applied to extract the predicted etch depth and sidewall angle.

We start with an initial simulator parameter vector $z_r^{(0)}$ and refine it iteratively using \(D_{\mathrm{fit}}^{(r)}\). Here, $z$ denotes the set of ViennaPS parameters to be calibrated, which will be detailed in Section~\ref{sec:para_space}. In the $t$-th iteration, we fix the current parameters $z_r^{(t)}$ and run ViennaPS for each input condition in the fitting set $D^{(r)}_{\text{fit}} = \{(x^{(r)}_i, y^{(r)}_i)\}_{i=1}^{N^{(r)}_{\text{fit}}}$. Since 
calibration is performed independently for each recipe $r$, the $i$-th input is defined as $x_i = (c^{(r)}_i, o^{(r)}_i, p_r)$, where $c^{(r)}_i$ and $o^{(r)}_i$ denote the cycle number and opening size, and $p_r$ is the fixed recipe setting.

For each sample in $D^{(r)}_{\text{fit}}$, we evaluate the current simulator parameter by running ViennaPS as $s_i^{(r,t)} = S(x^{(r)}_i; z_r^{(t)})$. We then apply the profile measurement operator $\phi(\cdot)$ to extract etch depth and sidewall angle from the simulated profile(according to Section~\ref{sec:Profile_Measure}). The measured simulator response $\phi(s_i^{(r,t)})$ is compared with the corresponding real observation $\phi(y^{(r)}_i)$, yielding a global error over $D^{(r)}_{\text{fit}}$. During the early search stage, the candidate parameters are mainly generated from local random exploration and LLM-guided updates(according to Section~\ref{sec:llm}). These candidates are then evaluated by ViennaPS simulator, forming the data collection loop shown in the blue dotted line box of Fig~\ref{fig:framework}.

All evaluated simulations are stored in a historical dataset. Once a sufficient number of simulation records has been accumulated, we train an XGBoost surrogate to approximate the forward response of ViennaPS and estimate the quality of unevaluated parameter candidates(according to Section~\ref{sec:xgboost}). The surrogate is then used to screen a larger candidate pool and propose additional high-potential parameters. Therefore, the parameter pool is progressively updated from
\begin{equation}
\mathcal{P}^{(t+1)}=\mathcal{P}_{\text{ram}}^{(t+1)}
\cup·
\mathcal{P}_{\text{LLM}}^{(t+1)}
\end{equation}
in the data collection stage to
\begin{equation}
\mathcal{P}^{(t+1)}=\mathcal{P}_{\text{ram}}^{(t+1)}
\cup
\mathcal{P}_{\text{LLM}}^{(t+1)}
\cup
\mathcal{P}_{\text{xgb}}^{(t+1)}
\end{equation}
after the surrogate becomes available. More specifically, in each iteration, the candidate parameters in each iteration consist of three components: (1) physics-guided candidates proposed by the LLM based on the current global error; (2) random local exploration candidates used to enhance parameter space coverage; (3) high-potential candidates selected based on predictions from the online surrogate. This hybrid candidate pool design balances physical priors, search efficiency, and exploration capability.

Finally, all candidates are evaluated by running ViennaPS over $D_{\text{fit}}$, and the one yielding the lowest global error is selected to update the parameters. It is important to note that the XGBoost surrogate is used only to assist candidate generation and preliminary screening. The final parameter update is always determined by ViennaPS evaluation rather than surrogate prediction. This process repeats until convergence or a stopping criterion is met, resulting in the identified parameters $z_r^*$.

\subsection{ViennaPS and Parameter Space}
\label{sec:para_space}

ViennaPS \cite{reiter2025viennaps} is an open-source framework for simulating topography evolution in semiconductor fabrication processes, especially etching and deposition. In this work, the Bosch process is implemented using a one-time ignition stage followed by repeated cycles of deposition, Etch A, and Etch B, as described in Section~\ref{bosch_process}. This process template is used throughout the calibration under a fixed recipe setting.

Although ViennaPS contains many physical and numerical parameters, in this work, we focus on the four main process stages in the simulation template: the initial ignition stage, the deposition stage, and the two etching stages. We assign separate flux-related parameters to each stage. These parameters control the supply of ion and neutral species arriving at the wafer surface, and are therefore selected as the identifiable simulator parameters. This constrained parameterization reduces the search dimension and also keeps the key physical factors that dominate Bosch profile evolution. Therefore, the calibration problem becomes more feasible under limited real experimental data.

Based on this design, we define the identifiable simulator parameter vector as

\begin{equation}
z = \bigl[\mathrm{Flux}_{ionI}, \mathrm{Flux}_{neuI}, \mathrm{Flux}_{ionD}, \mathrm{Flux}_{neuD}, \mathrm{Flux}_{ionE_A}, \mathrm{Flux}_{neuE_A}, \mathrm{Flux}_{ionE_B}, \mathrm{Flux}_{neuE_B}\bigr]^{\top}.
\end{equation}

Here, $\mathrm{Flux}_{ion}$ represents the ion flux arriving at the wafer surface, primarily corresponding to ion-assisted surface reactions, while $\mathrm{Flux}_{neu}$ is the neutral particle flux associated with chemical reaction channels. Consequently, modulating these flux parameters essentially regulates the supply intensity of effective reactants across different process stages. Specifically, these parameters are partitioned according to the distinct phases of the Bosch cycle. 
$\mathrm{Flux}_{ionI}$ and $\mathrm{Flux}_{neuI}$ correspond to the initial ignition phase, influencing initial opening formation and early profile evolution, $\mathrm{Flux}_{ionD}$ and $\mathrm{Flux}_{neuD}$ correspond to the deposition phase, and $\mathrm{Flux}_{ionE_A}$, $\mathrm{Flux}_{neuE_A}$,$ \mathrm{Flux}_{ionE_B}$ and $ \mathrm{Flux}_{neuE_B}$ correspond to the two etching steps within a period.

\subsection{Etch Profile Measurement}
\label{sec:Profile_Measure}

\begin{figure}[t]
    \centering
    \includegraphics[width=0.5\textwidth]{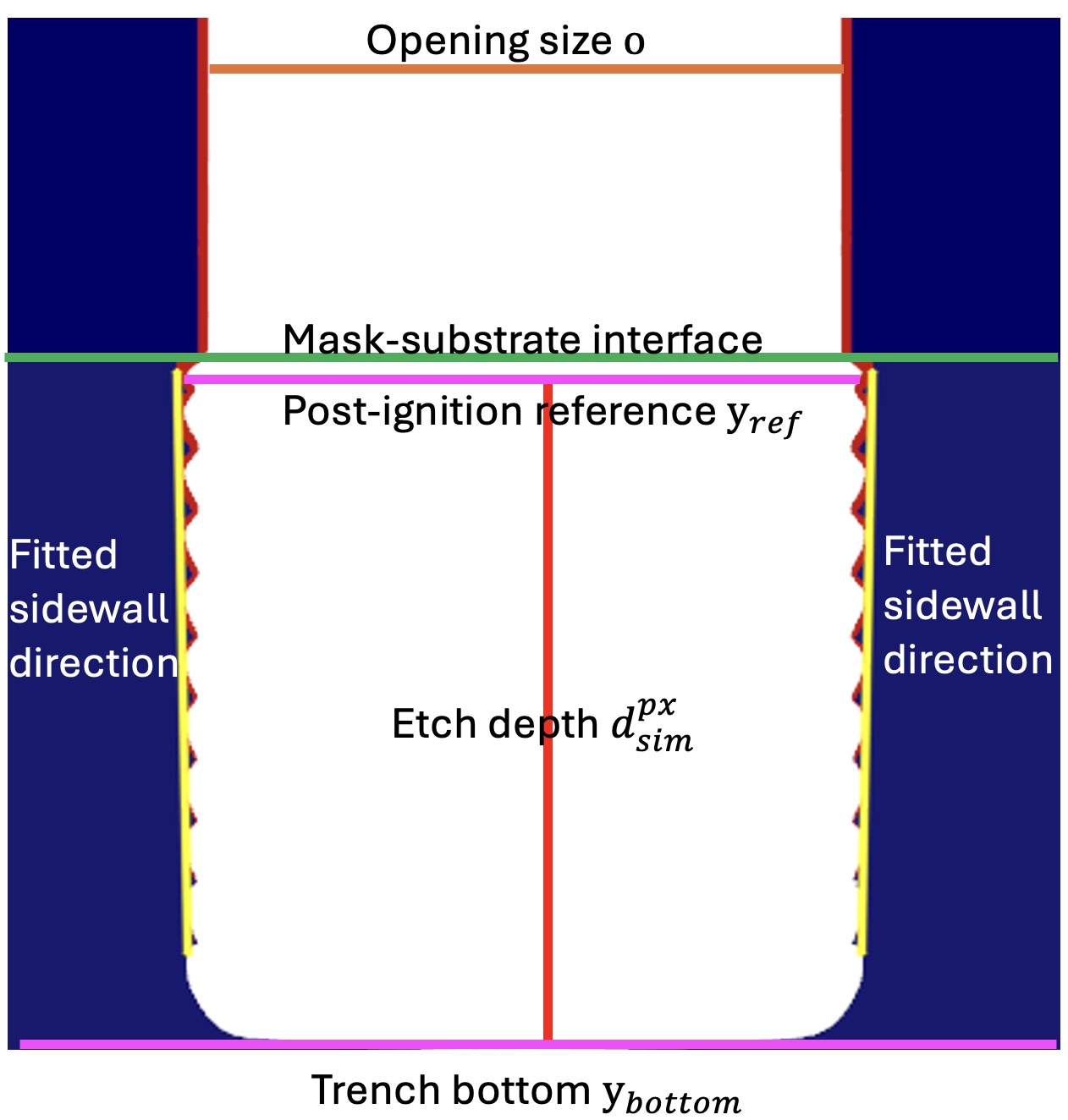}
    \caption{Etch Profile Measurement.}
    \label{sim_measure}
\end{figure}

In this section, we describe how etch depth and sidewall angle are automatically extracted from the simulated profiles. Etch depth and sidewall angle are estimated through classic computer vision, as shown in Fig.~\ref{sim_measure}. The orange horizontal line indicates the opening size \(o\), and the green line marks the mask-substrate interface. The upper magenta line denotes the post-ignition reference position \(y_{\mathrm{ref}}\), while the lower magenta line denotes the trench bottom \(y_{\mathrm{bottom}}\).

Because the ignition step is used only to form the initial opening, the reported etch depth excludes the profile generated during ignition. It is measured as the vertical pixel distance from the post-ignition reference position to the lowest point of the trench:
\begin{equation}
    d_{\mathrm{sim}}^{\mathrm{px}} = y_{\mathrm{bottom}} - y_{\mathrm{ref}} .
\end{equation}

When the physical opening size is available, this pixel distance is converted into micrometers using the pixel-to-length ratio.

To estimate the sidewall angle, we identify the widest positions of the first and last detected sidewall scallops on each side. The yellow fitted line
on each side connects these two positions and defines the corresponding sidewall direction. Let \(\theta\) denote the angle between the fitted sidewall direction and the vertical direction. A positive $\theta$ indicates inward tapering
(the trench narrows with depth), while a negative
$\theta$ indicates outward tapering. The sidewall angle is defined as
\begin{equation}
    \alpha = 90^\circ - \theta.
\end{equation}
The final sidewall angle is the average of the left and right sidewall angles.

We further define a validity label for each simulated profile. A profile is considered invalid if the measured depth is smaller than $1~\mu\mathrm{m}$ or the sidewall angle is smaller than $70^\circ$. Formally,
\begin{equation}
v(s)=
\begin{cases}
1, & d_{\mathrm{sim}} \geq 1~\mu\mathrm{m} \ \text{and} \ \alpha_{\mathrm{sim}} \geq 70^\circ,\\
0, & d_{\mathrm{sim}} < 1~\mu\mathrm{m} \ \text{or} \ \alpha_{\mathrm{sim}} < 70^\circ .
\end{cases}
\end{equation}
This validity check prevents clearly failed simulations from being selected as good parameter candidates only because of a misleading numerical error.


\subsection{LLM-Based Parameter Guidance}
\label{sec:llm}

For each recipe \(r\), the goal of calibration is to find a recipe-specific simulator parameter vector $z_r^*$ shared across all fitting conditions within that recipe $D^{(r)}_{\text{fit}}$. For a candidate parameter vector $z$, we need to evaluate its quality by applying it to every fitting condition $x_i \in D_{\text{fit}}$. In other words, ViennaPS is run with the same $z$ but different process inputs $x_i$, and the resulting simulated profiles are compared with the corresponding real measurements. Therefore, evaluating one candidate $z$ already requires multiple ViennaPS simulations, and evaluating many candidates leads to a large computational burden.

To search for $z^*$ efficiently, each iteration first generates a small candidate pool. These candidates can come from three sources: random local exploration, LLM-based updates, and XGBoost-based updates. Random exploration helps cover the parameter space, but it does not use physical feedback and can waste many simulations. XGBoost can learn from historical simulation records and propose promising candidates, but it is unreliable at the beginning of calibration because only a small number of records are available. Therefore, we introduce an LLM-based parameter guidance module to fill this gap. The LLM uses the global error, validity information, and Bosch process knowledge to suggest physically meaningful parameter updates before the surrogate becomes reliable.

Specifically, after each full evaluation, we summarize the error between the current simulation results and the real observations. For the $t$-th iteration, this summary can be written as

\begin{equation}
\mathcal{C}^{(t)}=
\left\{z^{(t)},\epsilon(z^{(t)}),e^{(t)},v^{(t)},\mathcal{M}\right\},
\end{equation}

where $e^{(t)}$ denotes the aggregated residual over $D_{\text{fit}}$, $v^{(t)}$ denotes the validity information, and $\mathcal{M}$ denotes the physical meaning of each flux parameter. Based on this summary, the LLM returns a small number of candidate parameter updates with numerical values and short physical explanations:

\begin{equation}
\mathcal{P}_{\text{LLM}}^{(t+1)}=\left\{(z_{\text{LLM},1}^{(t+1)},\text{info}_{\text{LLM},1}^{(t+1)}),\ldots,(z_{\text{LLM},K}^{(t+1)},\text{info}_{\text{LLM},K}^{(t+1)})\right\},
\end{equation}

where $K$ denotes the number of candidate parameter vectors proposed by the LLM.

This candidate generation is implemented through a structured prompt. The prompt is designed to make the LLM output controllable and easy to parse. It contains two parts. The first part describes the current calibration state, including the current parameter setting, the simulated depth and sidewall angle, the corresponding real measurements, and the residuals. The second part provides Bosch process knowledge. It describes the physical roles of ignition, etching, and deposition fluxes, and gives simple rules for interpreting depth and angle residuals. The task part instructs the LLM to analyze the depth error, sidewall angle error, and invalid simulation cases. Based on this analysis, the LLM suggests promising parameter update directions and identifies parameter regions that should be avoided. This allows the search to move toward physically reasonable regions. The output part requires the LLM to return recommended parameter updates in a structured format. Each output candidate contains explicit numerical flux values, a magnitude label, and a short physical reason.  These outputs are parsed as the LLM-based candidate pool and then evaluated by ViennaPS simulation. The complete prompt template is provided in Appendix~\ref{app:llm_prompt}.

In this way, the LLM complements both random exploration and XGBoost-based updates. It provides physically meaningful search directions in the early search stage, when there are not enough simulation records to train a reliable XGBoost surrogate. Compared with random exploration, the LLM can use the current error pattern to suggest more meaningful updates. It can also use validity information to avoid parameter changes that are likely to produce invalid profiles. In addition, each LLM proposal includes a short physical reason, which makes the search process more interpretable.

\subsection{XGBoost-Based Parameter Screening}
\label{sec:xgboost}

While both random exploration and LLM-based guidance can suggest new parameters, they do not fully make use of the historical simulation data collected during calibration. Random exploration helps cover more ground, but it ignores how the simulator has actually responded. The LLM can provide physically reasonable updates, but its  guidance relies mostly on process knowledge, which lacks a proper model for the behavior of ViennaPS under different process conditions.

As the calibration loop proceeds, each evaluated candidate produces useful simulation records, including the simulator parameter vector, measured depth, sidewall angle, and validity label. We train an XGBoost surrogate using these data to approximate the forward response of ViennaPS. The surrogate allows us to quickly estimate the potential quality of unevaluated candidates and screen a larger parameter pool before running full ViennaPS simulations.


Specifically, \(\mathcal{H}^{(t)}\) denotes the historical simulation dataset accumulated up to iteration \(t\):

\begin{equation}
\mathcal{H}^{(t)}=\left\{(x_i, z_j,\phi(S(x_i;z_j)), v_{ij})\right\}.
\end{equation}

Here, $i$ indexes the fitting sample in $D_{\text{fit}}$, and $j$ indexes the candidate simulator parameter vector that has been evaluated in previous iterations. Therefore, each record contains the process condition \(x_i\), the evaluated simulator parameter vector \(z_j\), the measured simulator measurements \(\phi(S(x_i;z_j))\), and the validity label \(v_{ij}\). The surrogate's input is a combination of the process condition and the simulator parameter vector:

\begin{equation}
q_{ij}=[c_i,o_i,z_j],
\end{equation}

where $c_i$ is the cycle number, $o_i$ is the opening size, and $z_j$ is the evaluated flux parameter vector. Based on this input, the XGBoost surrogate learns to predict how ViennaPS behaves. It predicts the simulated etch depth and sidewall angle for the process condition $x_i$ and parameter vector $z_j$:

\begin{equation}
\tilde{d}_{ij}=f_d(q_{ij}), \qquad \tilde{\alpha}_{ij}=f_\alpha(q_{ij}),   
\end{equation}

here, $f_d$ and $f_\alpha$ are two XGBoost regressors trained on the historical simulation records. 

The surrogate is trained with the following regression objective:

\begin{equation}
\min_{f_d,f_\alpha} \sum_{(i,j)\in \mathcal{H}^{(t)}}
\left[ \left(f_d(q_{ij})-d_{ij}^{\mathrm{sim}}
\right)^2 + \left( f_\alpha(q_{ij})-\alpha_{ij}^{\mathrm{sim}}
\right)^2 \right] + \Omega(f_d) + \Omega(f_\alpha),
\end{equation}

where \(\Omega(\cdot)\) denotes the regularization term of the XGBoost model.

In addition, we train a validity classifier based on valid and invalid simulation records:
\begin{equation}
\hat{v}_{ij}=f_v(q_{ij}),  
\end{equation}
where $\hat{v}_{ij}$ estimates whether the candidate parameter is likely to produce a valid simulated profile. The classifier is trained with a binary cross-entropy objective:
\begin{equation}
\min_{f_v} \sum_{(i,j)\in \mathcal{H}^{(t)}} -
\left[v_{ij}\log f_v(q_{ij})+(1-v_{ij})\log(1-f_v(q_{ij}))\right],   
\end{equation}
where \(v_{ij}=1\) indicates a valid simulated profile and \(v_{ij}=0\) indicates an invalid one.

For an unevaluated candidate $z$, the surrogate predicts depth and sidewall angle for each fitting condition. These predictions are compared with the real measurements to compute a surrogate score:
\begin{equation}
\tilde{\epsilon}_{\text{xgb}}(z) = \frac{1}{N_{\text{fit}}} \sum_{(x_i,y_i)\in D_{\text{fit}}} \mathcal{L} \left( [\tilde{d}(x_i,z),\tilde{\alpha}(x_i,z)]^\top, \phi(y_i) \right).    
\end{equation}

Candidates with lower predicted error and acceptable validity probability are selected as $\mathcal{P}_{\text{xgb}}^{(t+1)}.$ Then these selected candidates are added to the candidate pool and are still evaluated by the ViennaPS simulation over $D_{\text{fit}}$. 

\section{Experiments}
\label{sec:Experiments}
\subsection{Experimental Setup and Evaluation Metrics}
\label{sec:Exp_setup_metric}
\paragraph{\textbf{Real Experiment Data}}

Table~\ref{tab:real_data_coverage} shows the coverage of the real Bosch etching samples for the two recipe settings in terms of cycle number, opening size, measured etch depth and sidewall angle. 
For each recipe $r$, the real dataset is divided into two non-overlapping subsets:$D_{\text{fit}}^{(r)}$ and $D_{\text{test}}^{(r)}$. The fitting set \(D_{\text{fit}}^{(r)}\) is used to identify the recipe-specific simulator parameter vector \(z^*_r\). 
The test set \(D_{\text{test}}^{(r)}\) is used to validate the efficacy of the framework. The parameter identification stage for each recipe strictly excludes any test samples. The calibration is performed independently for the two recipes, yielding a separate simulator parameter vector \({z}_r^*\) for each recipe.

Appendix~\ref{app:recipe_sample} provides the complete list of real samples for Recipe 1. It includes the measured values and their assignment to the fitting and test sets. It also provides the detailed process parameters for both Recipe 1 and Recipe 2.

\begin{table}[htbp]
\centering
\caption{Real Bosch etching samples used in this work. 
For each recipe, the cycle number, GDS design, and opening size vary across samples.}
\label{tab:real_data_coverage}

\small
\setlength{\tabcolsep}{4pt}

\begin{tabularx}{\textwidth}{
    c
    c
    c
    >{\centering\arraybackslash}X
    c
    c
}
\toprule
Cycle & No. of samples & GDS design &
Opening sizes ($\mu$m) &
Depth range ($\mu$m) &
Angle range ($^\circ$) \\
\midrule

\multicolumn{6}{c}{\textbf{Recipe 1}} \\
\midrule
10 & 3 & 6,10,10 & 6.20, 10.60, 11.00
& 7.05--7.84 & 88.74--90.91 \\

20 & 3 & 7,8,9 & 7.589, 8.344, 9.263
& 14.152--14.790 & 89.73--91.49 \\

30 & 4 & 7,8,9,9 & 7.15, 8.12, 9.10, 9.858
& 19.49--20.475 & 89.94--90.31 \\

40 & 4 & 7,8,9,10 & 7.552, 8.256, 9.116, 10.193
& 23.95--25.732 & 90.54--91.64 \\

70 & 6 & 6,7,8,9,10,15 & 6.95, 7.64, 8.72, 9.80, 10.81, 15.82
& 38.86--46.57 & 89.60--90.36 \\

\midrule
\multicolumn{6}{c}{\textbf{Recipe 2}} \\
\midrule
10 & 5 & 3,3,4,5,15 & 3.02, 3.48, 4, 5.46, 14.69 & 5.43--6.29 & 89.31--90.51 \\
20 & 5 & 3,4,5,8,10 & 3.75, 4.84, 5.25, 8.25, 10.75 & 9.93--11.71 & 89.46--91.68 \\
30 & 6 & 6,7,8,9,10,15  & 5.98,6.93, 8.13,9.27, 11.2,16.06 & 13.25--15.88 & 88.89–90.58 \\
70 & 5 & 10,15,15,15,20 & 8.22,12.12, 12.87,13.47, 18.78 & 27.69--31.13 & 88.48–92.03 \\
\bottomrule
\end{tabularx}
\end{table}

\begin{figure}[!htbp]
    \centering

    \begin{minipage}[b]{0.18\columnwidth}
        \centering
        \includegraphics[width=\linewidth]{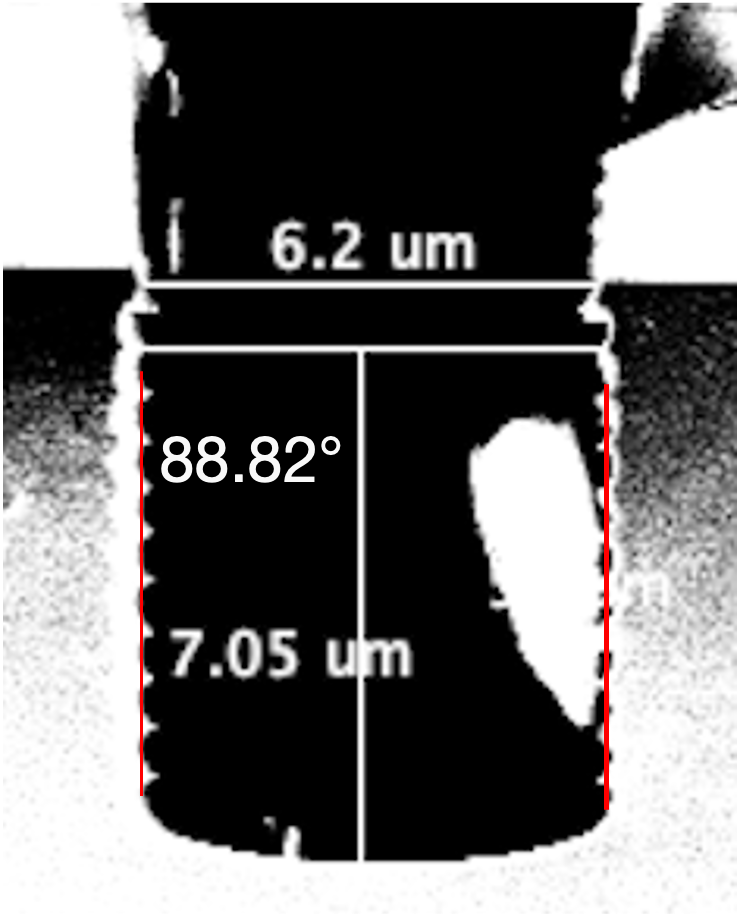}
        \par\smallskip
        \textbf{(a) 10 cycles}
    \end{minipage}
    \hspace{0.02\columnwidth}
    \begin{minipage}[b]{0.18\columnwidth}
        \centering
        \includegraphics[width=\linewidth]{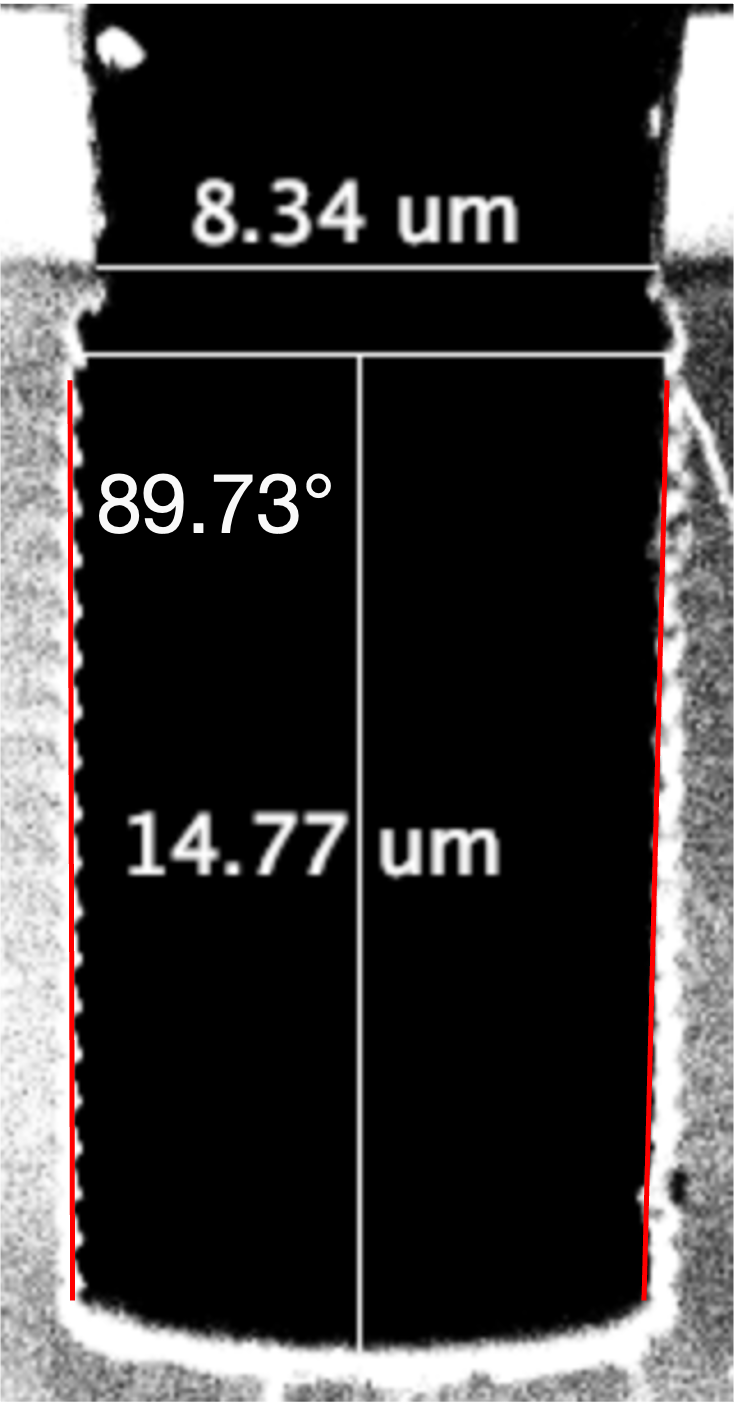}
        \par\smallskip
        \textbf{(b) 20 cycles}
    \end{minipage}
    \hspace{0.02\columnwidth}
    \begin{minipage}[b]{0.18\columnwidth}
        \centering
        \includegraphics[width=\linewidth]{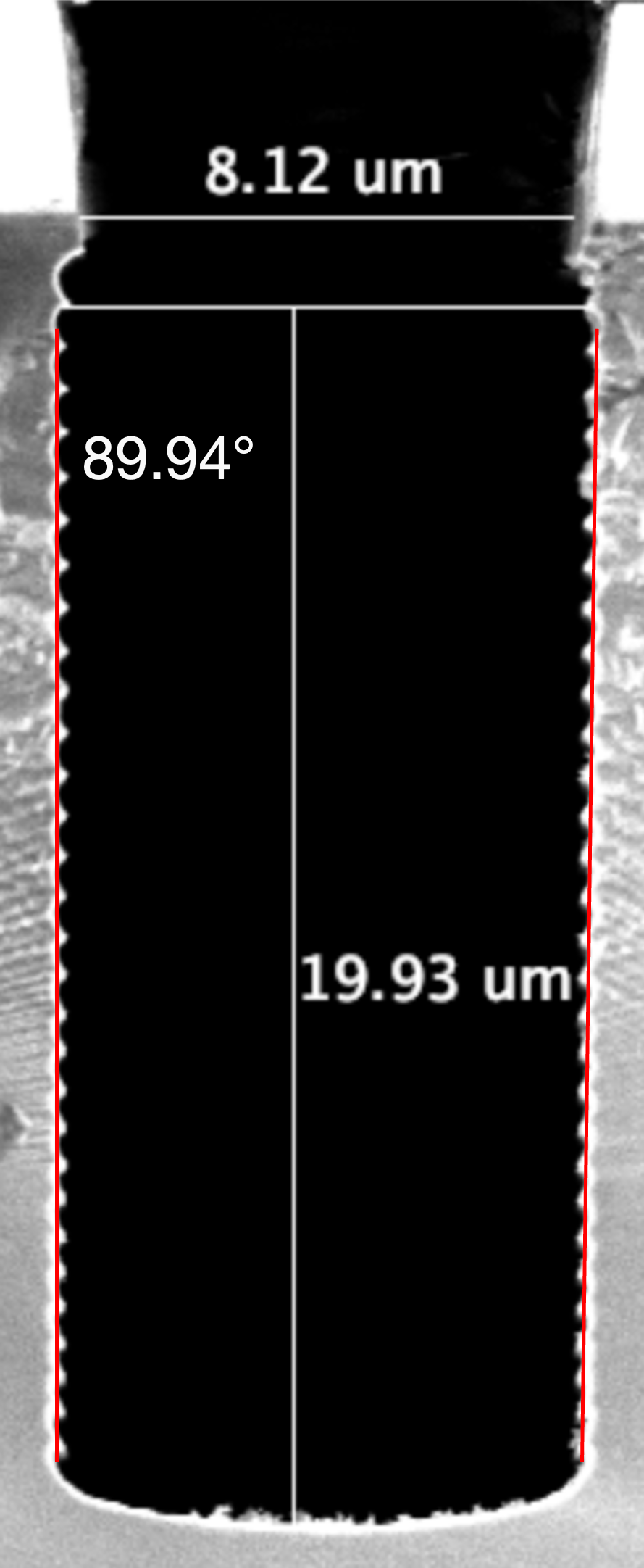}
        \par\smallskip
        \textbf{(c) 30 cycles}
    \end{minipage}

    \caption{Representative real XSEM profiles used in this work. The top horizontal line indicates the opening size, the vertical line indicates the etch depth, and the red dashed sidewall lines are fitted by connecting the first and last scallop positions on each side, from which the sidewall angle is measured.}
    \label{fig:real_xsem}
\end{figure}

The real etching profiles are manually measured using ImageJ to obtain the etch depth and sidewall angle. The opening size is measured as the horizontal distance between the two mask edge positions at the top opening. The etch depth is measured from the end of the first scallop to the trench bottom, so that the initial opening formed during ignition is excluded, consistent with the simulation measurement. The sidewall angle is measured by fitting the left and right sidewall directions between representative scallop positions and averaging the two angles. Figure~\ref{fig:real_xsem} shows representative annotated real XSEM profiles illustrating these measurements. These measurements serve as the ground-truth labels for simulator calibration and final prediction evaluation.

\paragraph{\textbf{ViennaPS Setup}}

All simulations are performed using the ViennaPS Bosch ray-tracing process simulator. 
The process is implemented using a one-time ignition
stage followed by repeated cycles of deposition, Etch A, and Etch B.

For clarity, we describe the ViennaPS setup using Recipe 1 as an example. The same calibration procedure is independently applied to Recipe 2. For the \(i\)-th sample in Recipe 1, the simulator input is
\(x_i=(c_i,o_i,p_1),\)
where \(c_i\) and \(o_i\) are the sample-specific cycle number and opening size, respectively, and \(p_1\) is the fixed experimental recipe reported in Table~\ref{tab:real_recipe} Recipe 1. Calibration starts from the initial simulator parameter vector \(z^{(0)}_1\), whose eight flux values are listed in Table~\ref{tab:initial_flux_parameters}.

During calibration, only the eight flux-related parameters defined in Section~\ref{sec:para_space} are updated. Geometry parameters, step times, sticking probabilities, effective reaction coefficients and other ViennaPS settings are all kept fixed. The detailed fixed ViennaPS parameters are provided in Appendix~\ref{app:Vienna_recipe}. For each candidate parameter vector \(z\), ViennaPS generates a simulated etch profile. 
The profile measurement procedure in Section~\ref{sec:Profile_Measure} is then used to extract etch depth and sidewall angle.

\paragraph{\textbf{Evaluation Metrics}}
We evaluate the proposed calibration framework from three aspects: calibration accuracy, prediction accuracy, and ablation study. In each recipe,
calibration accuracy measures how well the identified parameter vector \(z_r^*\) aligns ViennaPS with real measurements on the fitting set \(D^{(r)}_{\text{fit}}\). 
Prediction accuracy measures whether the identified parameter vector \(z_r^*\) can be directly used for held-out process conditions within the same recipe. 
For each test sample \(x^{(r)}_j \in D^{(r)}_{\text{test}}\), we run ViennaPS once with the fixed calibrated parameter vector $z_r^*$, obtaining the predicted simulated profile \(\hat{s}^{(r)}_j=S(x^{(r)}_j;z_r^*)\). Then we extract etch depth and sidewall angle as \(\phi(\hat{s}^{(r)}_j)\), and compare them with the corresponding real measurement \(\phi(y^{(r)}_j)\). 
Search efficiency measures how many calibration iterations and ViennaPS evaluations are required to identify \(z_r^*\).

For both calibration and prediction accuracy, we report the same accuracy metrics, including depth mean absolute error, sidewall-angle mean absolute error, relative depth error, and normalized global error. 
The depth MAE is defined as:
\begin{equation}
\mathrm{MAE}_d = \frac{1}{N} \sum_{i=1}^{N}
\left|d_i^{\mathrm{sim}}-d_i^{\mathrm{real}}\right|.
\end{equation}

The sidewall-angle MAE is defined as:
\begin{equation}
\mathrm{MAE}_{\alpha} = \frac{1}{N} \sum_{i=1}^{N}
\left|\alpha_i^{\mathrm{sim}}-\alpha_i^{\mathrm{real}}\right|.
\end{equation}
Since etch depth may vary across different process conditions, we also report the relative depth error:
\begin{equation}
\mathrm{RE}_d = \frac{1}{N} \sum_{i=1}^{N}
\frac{\left|d_i^{\mathrm{sim}}-d_i^{\mathrm{real}}\right|}{d_i^{\mathrm{real}}}\times 100\% .
\end{equation}
To jointly evaluate etch depth and sidewall angle, we report a normalized global error:
\begin{equation}
\epsilon(z;D)=
\frac{1}{N}\sum_{(x_i,y_i)\in D}
\left[
w_d
\frac{|d_i^{\mathrm{sim}}-d_i^{\mathrm{real}}|}{\tau_d d_i^{\mathrm{real}}}
+
w_\alpha
\frac{|\alpha_i^{\mathrm{sim}}-\alpha_i^{\mathrm{real}}|}{\tau_\alpha}
\right],
\end{equation}

where \(D\) can be either \(D_{\text{fit}}\) or \(D_{\text{test}}\). 
The first term measures the relative depth error, which is more appropriate for comparing samples across different cycle numbers as etch depth increases with cycle number. The second term measures the absolute sidewall-angle error normalized by an angular tolerance. In this work, we set $\tau_d = 10\%$, $\tau_\alpha = 2^\circ$, and $w_d = w_\alpha = 0.5$, so that a $10\%$ depth error and a $2^\circ$ sidewall-angle error contribute equally to the overall error.

For ablation, we evaluate whether each candidate generation strategy can find a low-error simulator parameter vector with fewer ViennaPS evaluations. The main quantity used for comparison is the best calibration error found during the search. We define the best global error found up to iteration \(t\) as
\begin{equation}
\epsilon_{\mathrm{best}}^{(t)}
=
\min_{z \in \mathcal{Z}_{\mathrm{eval}}^{(t)}}
\epsilon(z;D{\mathrm{fit}}).
\label{global_error}
\end{equation}

Here, \(\mathcal{Z}_{\mathrm{eval}}^{(t)}\) is the set of all simulator parameter vectors that have been evaluated by ViennaPS up to iteration $t$.

\subsection{Calibration Accuracy}
\label{sec:calibration_accuracy}

In this section, we evaluate the calibration accuracy for both recipes.
Recipe 1 and Recipe 2 are calibrated independently, resulting in \(z_1^*\) and \(z_2^*\), respectively.
Each calibrated parameter vector is evaluated on its corresponding fitting set, \(D_{\mathrm{fit}}^{(1)}\) or \(D_{\mathrm{fit}}^{(2)}\), by comparing the ViennaPS simulations with the real XSEM measurements.

In addition to the metrics defined in Section~\ref{sec:Exp_setup_metric}, we report the relative improvement over the default ViennaPS parameter vector \(z_r^{(0)}\). For depth and sidewall angle, the improvements are computed as
\begin{equation}
\mathrm{Improvement}^{(r)}_d =
\frac{
\mathrm{MAE}_d(z_r^{(0)})-\mathrm{MAE}_d(z_r^*)
}{
\mathrm{MAE}_d(z_r^{(0)})
}
\times 100\% ,
\end{equation}

\begin{equation}
\mathrm{Improvement}^{(r)}_{\alpha} =
\frac{
\mathrm{MAE}_{\alpha}(z_r^{(0)})-\mathrm{MAE}_{\alpha}(z_r^*)
}{
\mathrm{MAE}_{\alpha}(z_r^{(0)})
}
\times 100\% .
\end{equation}
Here, the improvement measures how much the calibrated parameter vector reduces the fitting error compared with the uncalibrated simulator.

Since the original ViennaPS configuration contains only one etching step and no ignition step, for a fair comparison, the default baseline is not evaluated using the original single-etch ViennaPS process directly. Therefore, we construct a converted default parameter vector \(z_r^{(0)}\) under the same current process template used by \(z_r^*\), which includes an ignition step and two etching steps. All geometry settings, process times, sticking coefficients, effective reaction coefficients, the Etch A and Etch B sequence are kept identical to the calibrated simulator. Only the flux vector is initialized from the original ViennaPS default values. The deposition fluxes and the Etch A fluxes are copied from the original default configuration, while the ignition and Etch B fluxes are scaled according to the corresponding recipe flow rate and ICP power.

Table~\ref{tab:calibration_summary} reports the quantitative calibration results for both recipes. For Recipe 1, compared with the default parameter vector $z_1^{(0)}$, the calibrated $z^*_1$ reduces the depth MAE from \(18.063~\mu\mathrm{m}\) to $0.410~\mu\mathrm{m}$, corresponding to a \(97.729\%\) improvement. The relative depth error also decreases from \(84.858\%\) to \(1.970\%\), showing that the calibrated simulator can reproduce the etch rate behavior across different fitting conditions. For the sidewall angle MAE decreases from \(5.367^\circ\) to $1.207^\circ$, with a \(77.512\%\) improvement. The normalized global error is reduced from 5.585 to 0.400, indicating that the calibrated parameter vector \(z_1^*\) improves both etch depth and sidewall-shape agreement.
For Recipe 2, the independently identified parameter vector \(z_2^*\) similarly reduces the calibration error from $5.163$ to $0.237$. These results indicate that the proposed framework can identify a recipe-specific simulator parameter vector that aligns ViennaPS with real XSEM measurements under different Bosch recipe settings.

\begin{table}[t]
\centering
\caption{Calibration accuracy on two recipes.
The improvement is computed relative to the default ViennaPS parameter vector
$z^{(0)}$ for each recipe.}
\label{tab:calibration_summary}
\begin{tabular}{llcccccc}
\toprule
Recipe & Parameter vector &
MAE$_d$ ($\mu$m) &
RE$_d$ (\%) &
MAE$_\alpha$ ($^\circ$) &
$\epsilon$ &
Improvement$_d$ (\%) &
Improvement$_\alpha$ (\%) \\
\midrule

Recipe 1
& Default $z_1^{(0)}$
& 18.063 & 84.858 & 5.367 & 5.585 & -- & -- \\

& Calibrated $z_1^*$
& 0.410 & 1.970 & 1.207 & 0.400 & 97.729 & 77.512 \\

\midrule

Recipe 2
& Default $z_2^{(0)}$
& 11.308 & 81.93  & 4.265 & 5.163 & -- & -- \\
& Calibrated $z_2^*$
& 0.079 & 0.59 &  0.830 & 0.237 & 99.301 & 80.54
 \\

\bottomrule
\end{tabular}
\end{table}

\begin{figure*}[t]
    \centering
    \setlength{\tabcolsep}{2pt}

    \begin{minipage}[t]{0.32\textwidth}
        \centering\small
        \begin{tabular}{cc}
            \textbf{Real XSEM} & \textbf{\shortstack{Simulation}} \\
            \includegraphics[width=0.48\linewidth,height=4.0cm,keepaspectratio]{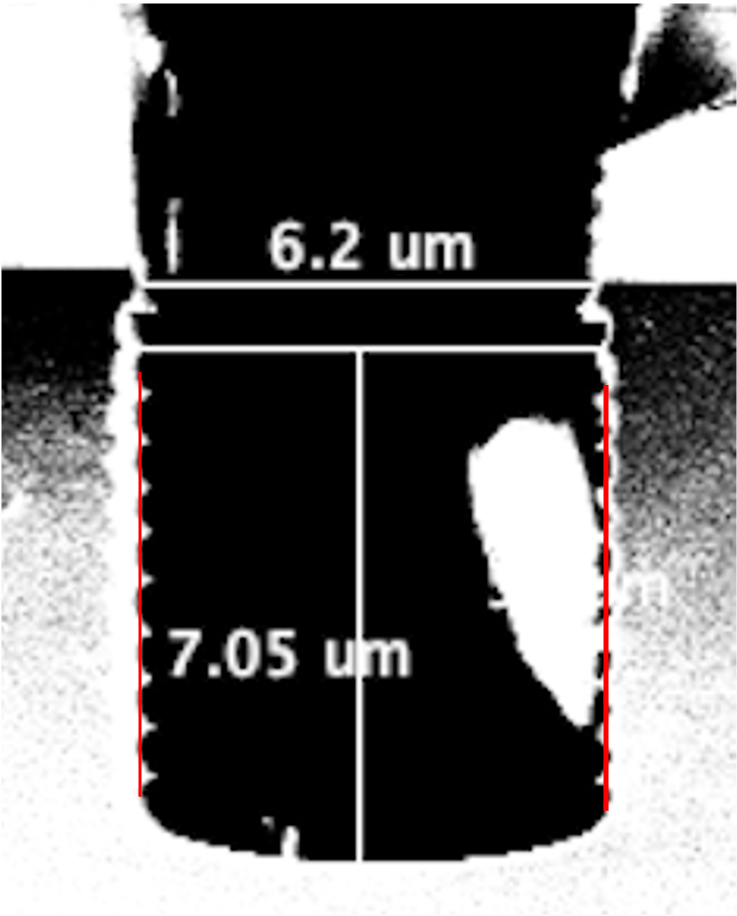} &
            \includegraphics[width=0.48\linewidth,height=4.0cm,keepaspectratio]{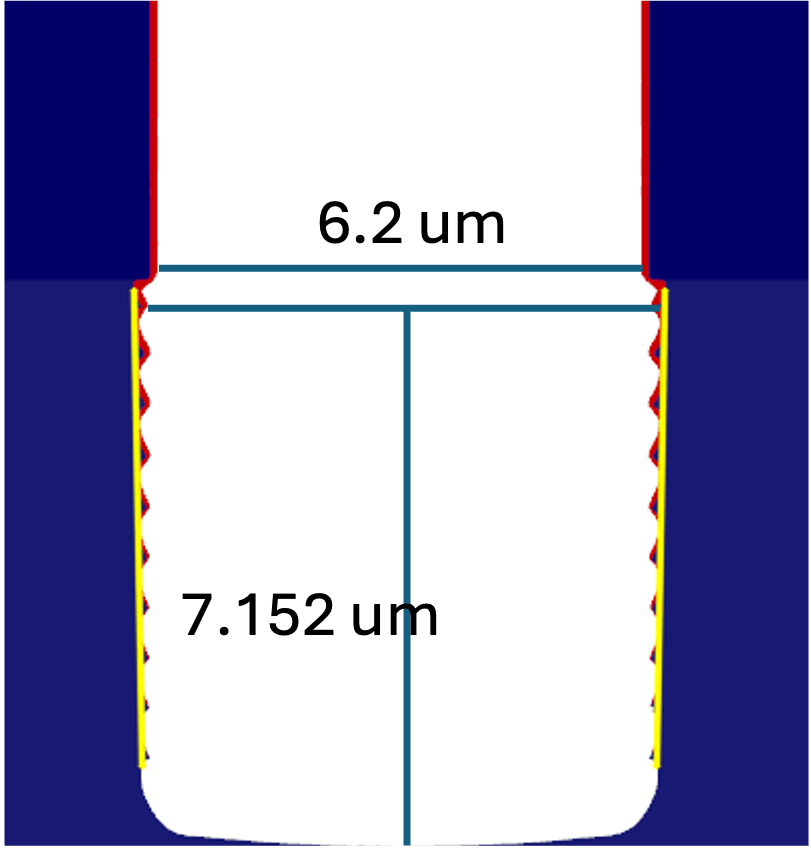}
        \end{tabular}
        \par\smallskip\textbf{(a)} Shallow etching: $c_a=10$, $o_a=6.20\,\mu\mathrm{m}$
        
        \label{fig:calib_profile_shallow}
    \end{minipage}
    \hfill
    \begin{minipage}[t]{0.32\textwidth}
        \centering\small
        \begin{tabular}{cc}
            \textbf{Real XSEM} & \textbf{\shortstack{Simulation}} \\
            \includegraphics[width=0.48\linewidth,height=4.0cm,keepaspectratio]{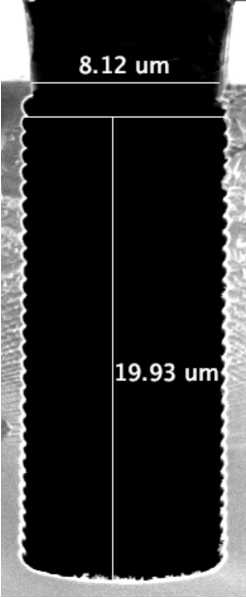} &
            \includegraphics[width=0.48\linewidth,height=4.0cm,keepaspectratio]{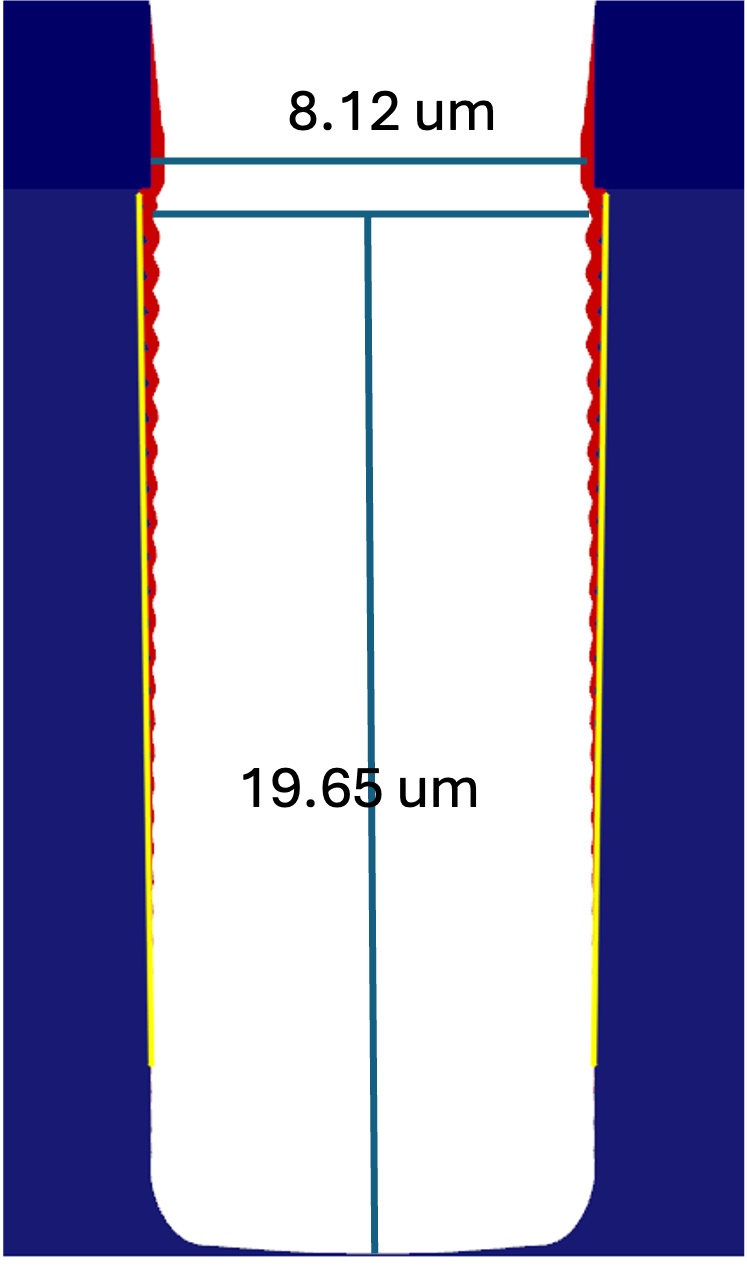}
        \end{tabular}
        \par\smallskip\textbf{(b)} Intermediate etching: $c_b=30$, $o_b=8.12\,\mu\mathrm{m}$
        \label{fig:calib_profile_intermediate}
    \end{minipage}
    \hfill
    \begin{minipage}[t]{0.32\textwidth}
        \centering\small
        \begin{tabular}{cc}
            \textbf{Real XSEM} & \textbf{\shortstack{Simulation}} \\
            \includegraphics[width=0.48\linewidth,height=4.0cm,keepaspectratio]{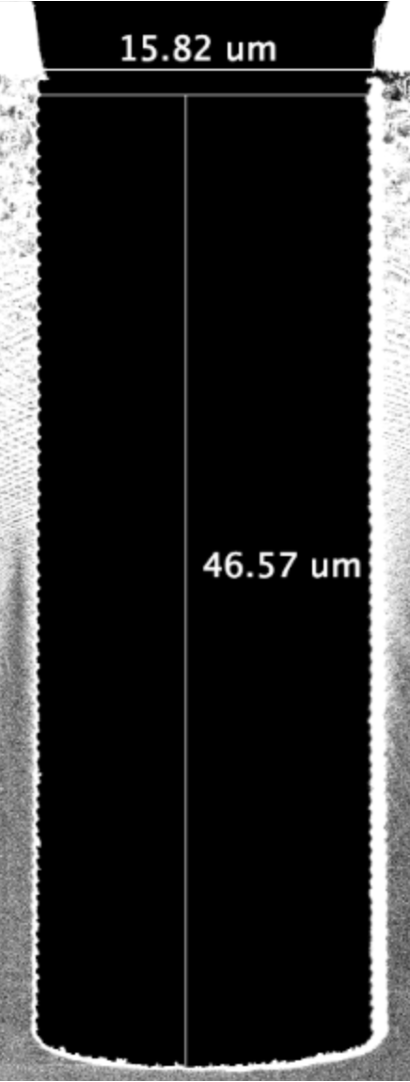} &
            \includegraphics[width=0.48\linewidth,height=4.0cm,keepaspectratio]{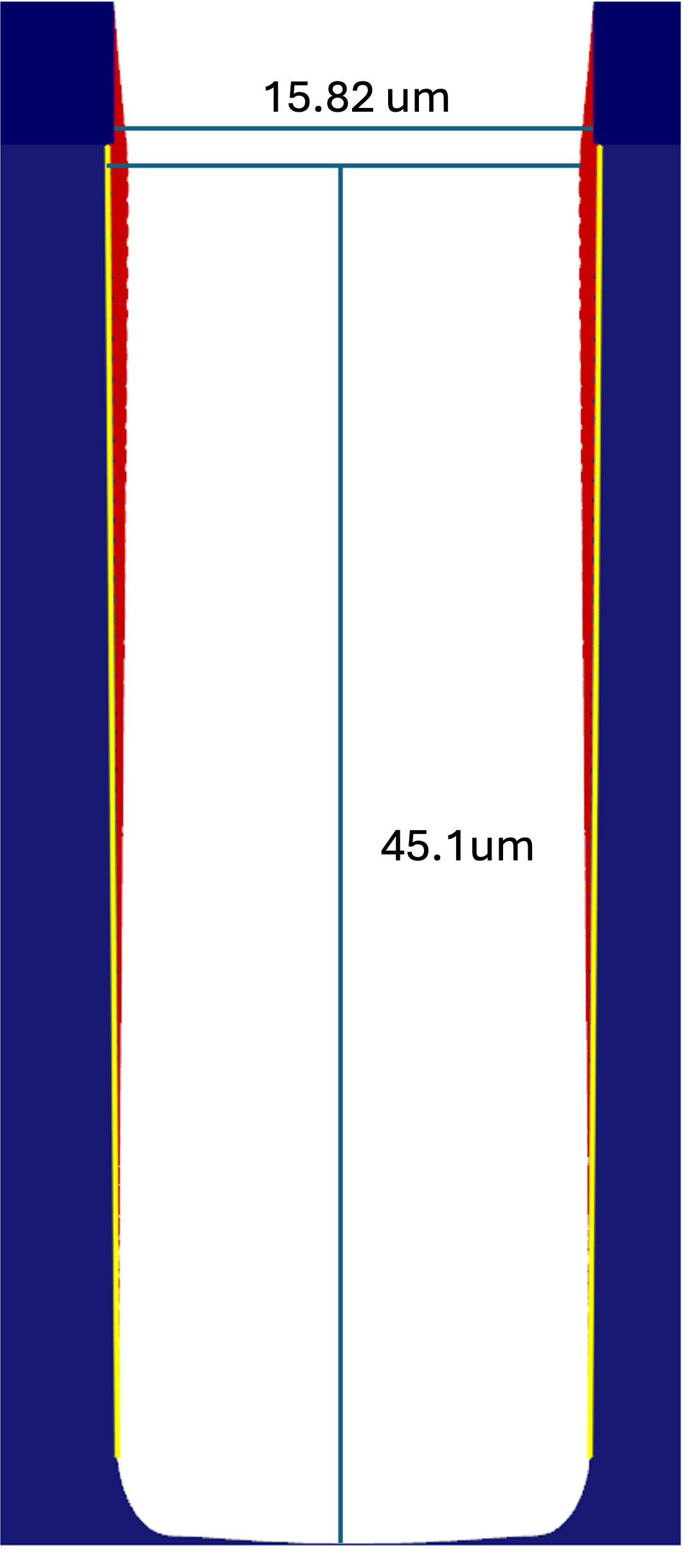}
        \end{tabular}
        \par\smallskip\textbf{(c)} Deep etching: $c_c=70$, $o_c=15.82\,\mu\mathrm{m}$
        \label{fig:calib_profile_deep}
    \end{minipage}

    \caption{Three representative fitted profiles for Recipe 1. Covering shallow, intermediate, and deep etching. Each paired comparison shows the real XSEM profile on the left and the corresponding calibrated ViennaPS simulation on the right. The examples correspond to different cycle numbers and opening sizes, demonstrating that the same calibrated parameter vector \(z_1^*\) can reproduce real profiles across multiple fitting conditions.}
    \label{fig:calib_profiles}
\end{figure*}

For qualitative visualization, figure~\ref{fig:calib_profiles} shows three representative fitted profiles from Recipe 1. The three examples are selected from \(D_{\mathrm{fit}}^{(1)}\) and correspond to shallow, intermediate, and deep etching conditions. Their process inputs are \(x_a^{(1)}=(c_a,o_a,p_1)\), \(x_b^{(1)}=(c_b,o_b,p_1)\), and \(x_c^{(1)}=(c_c,o_c,p_1)\), respectively. Despite the different cycle numbers and opening sizes, the same calibrated parameter vector \(z_1^*\) produces profiles that are well aligned with the corresponding XSEM measurements.

\begin{figure}[t]
\centering
\includegraphics[width=\linewidth]
{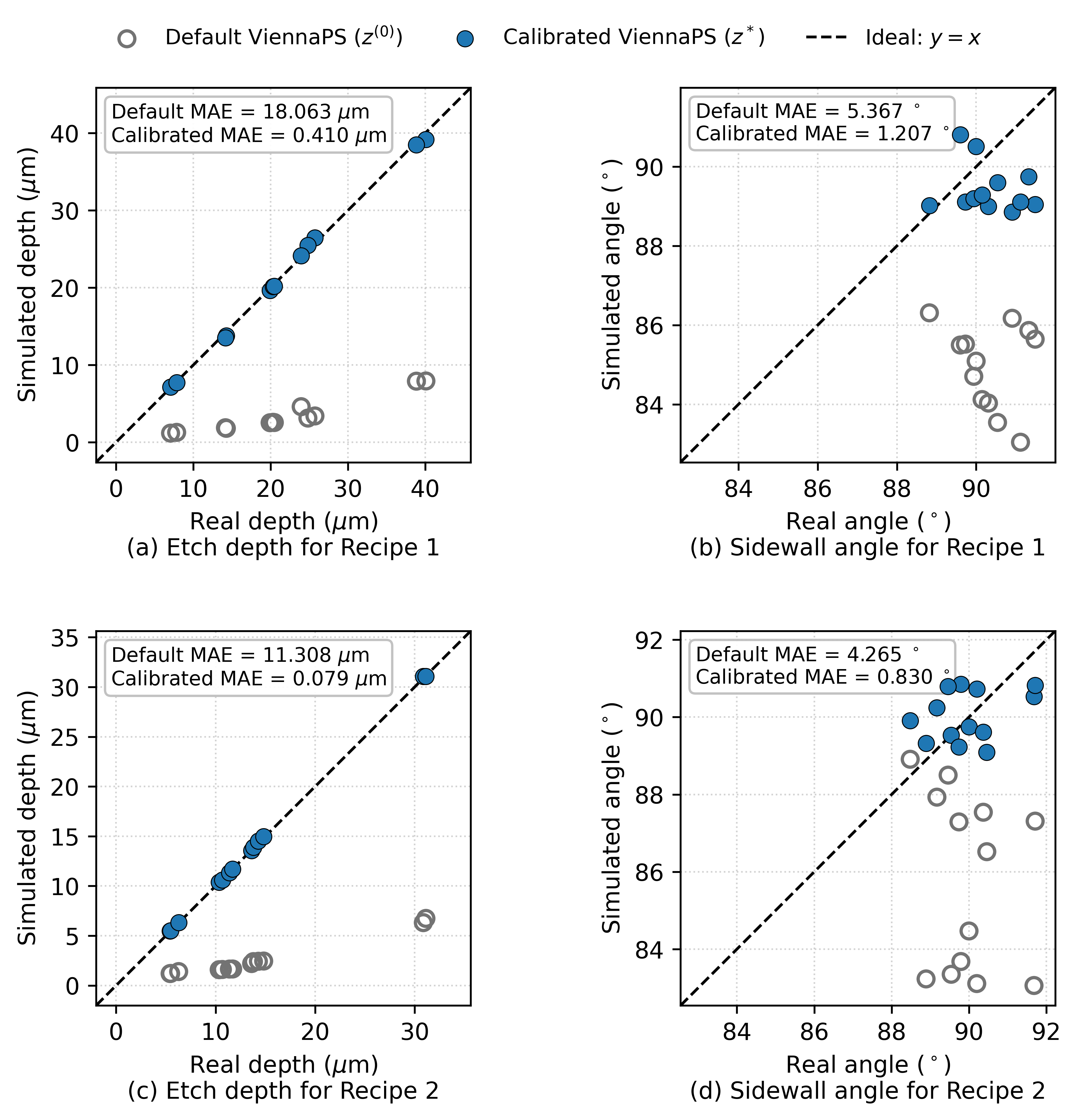}

\caption{
Real measurements versus simulated measurements on the fitting sets of
Recipe 1 and Recipe 2.
The dashed line denotes the ideal \(y=x\) agreement.
Gray circles indicate the default ViennaPS results, and blue circles indicate
the calibrated ViennaPS results.
(a) Etch depth for Recipe 1.
(b) Sidewall angle for Recipe 1.
(c) Etch depth for Recipe 2.
(d) Sidewall angle for Recipe 2.
}
\label{fig:calib_scatter}
\end{figure}

To examine the fitting behavior over all samples in the fitting sets of both recipes, Figure~\ref{fig:calib_scatter} compares the real measurements with the simulated measurements before and after calibration. For both etch depth and sidewall angle, the calibrated results are closer to the ideal \(y=x\) line than the default ViennaPS results. This indicates that calibration improves the agreement between ViennaPS and real XSEM measurements across the full fitting set.

Overall, the quantitative metrics, representative profile comparisons, and real-simulation scatter plots consistently show that the calibrated parameter vector \(z^*_1\) and \(z^*_2\) align ViennaPS with the real Bosch etching measurements for their corresponding recipes.

\subsection{Prediction Accuracy}
\label{sec:prediction_accuracy}

This section evaluates whether the calibrated parameter vector $z_r^*$ can be directly used to predict etch depth and sidewall angle under held-out process conditions within their corresponding recipes. Different from Section~\ref{sec:calibration_accuracy}, all samples in $D^{(r)}_{\mathrm{test}}$ are excluded from parameter identification. For each test sample, we fix $z_r^*$, run ViennaPS, and compare the extracted etch depth and sidewall angle against the corresponding real XSEM measurements.

Table~\ref{tab:prediction_summary} shows the quantitative prediction accuracy on the held-out test sets of both recipes. For Recipe \(r\), each test input has the form \(x^{(r)}_j=(c^{(r)}_j,o^{(r)}_j,p_r),\) where \(p_r\) is the fixed recipe shared by all samples within Recipe \(r\), whereas the cycle number \(c_j\) and opening size \(o_j\) vary across the test samples.
For Recipe 1, compared with the default parameter vector \(z_1^{(0)}\), the calibrated parameter vector \(z_1^*\) reduces the depth MAE from \(25.682~\mu\mathrm{m}\) to \(0.782~\mu\mathrm{m}\) and decreases the relative depth error from \(85.443\%\) to \(2.546\%\). The sidewall angle MAE is also reduced from \(6.103^\circ\) to \(0.852^\circ\). The normalized global error decreases from 5.798 to 0.340, indicating that the calibrated simulator improves both etch depth prediction and sidewall angle prediction on test samples. 
For Recipe 2, \(z_2^*\) similarly reduces the normalized global error from \(5.198\) to \(0.249\).
Since all samples in \(D^{(r)}_{\mathrm{test}}\) are excluded from parameter identification, these results show that the identified recipe-specific parameter vector \(z_r^*\) is not only fitted to \(D^{(r)}_{\mathrm{fit}}\), but can also generalize to held-out process conditions within the same recipe $r$.

\begin{table}[t]
\centering
\caption{Prediction accuracy on the held-out test sets for Recipe 1 and Recipe 2.}
\label{tab:prediction_summary}
\begin{tabular}{llcccc}
\toprule
Recipe
& Parameter vector 
& \(\mathrm{MAE}_d\) 
& \(\mathrm{RE}_d\)
& \(\mathrm{MAE}_{\alpha}\)
& \(\epsilon(z;D_{\mathrm{test}})\)
\\
&
& \((\mu\mathrm{m})\)
& \((\%)\)
& \((^\circ)\)
&
\\
\midrule

\multirow{2}{*}{Recipe 1}
& Default \(z_1^{(0)}\)
& 25.682 & 85.443 & 6.103 & 5.798 \\
& Calibrated \(z_1^*\)
& 0.782 & 2.546 & 0.852 & 0.340 \\

\midrule

\multirow{2}{*}{Recipe 2}
& Default \(z_2^{(0)}\)
& 14.345 & 81.71 & 4.451 & 5.198 \\
& Calibrated \(z_2^*\)
& 0.146 & 0.82 & 0.831 & 0.249 \\

\bottomrule
\end{tabular}
\end{table}

\begin{figure*}[t]
    \centering
    \setlength{\tabcolsep}{2pt}

    \begin{minipage}[t]{0.32\textwidth}
        \centering\small
        \begin{tabular}{cc}
            \textbf{Real XSEM} & \textbf{\shortstack{Calibrated\\Simulation}} \\
            \includegraphics[width=0.48\linewidth,height=4.0cm,keepaspectratio]{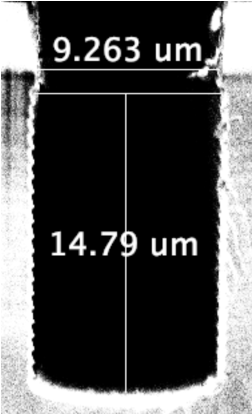} &
            \includegraphics[width=0.48\linewidth,height=4.0cm,keepaspectratio]{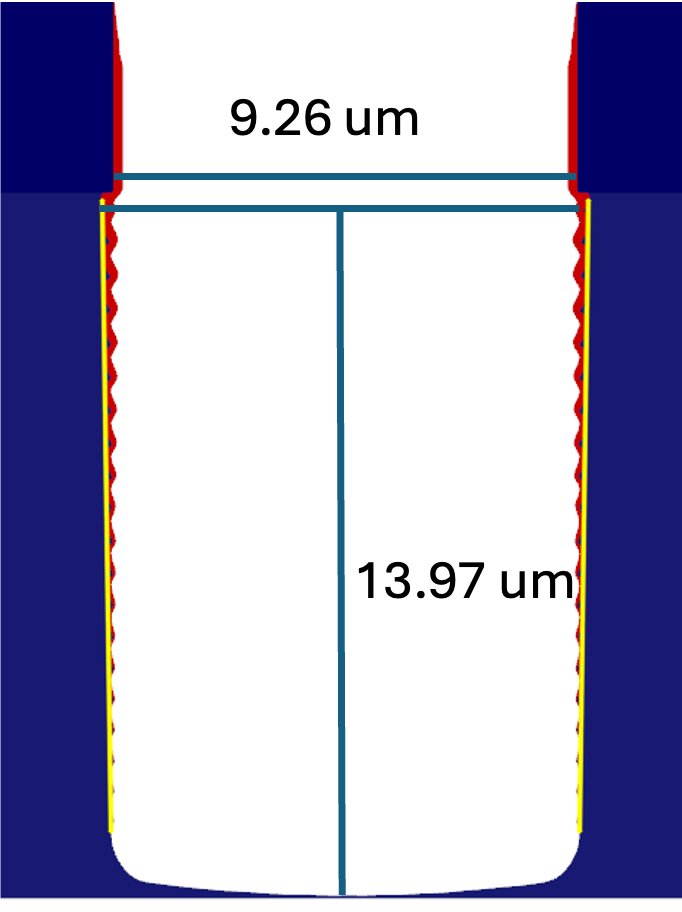}
        \end{tabular}
        \par\smallskip\textbf{(a)} Shallow etching:
$c_a^{test}=20$, $o_a^{test}=9.26\,\mu\mathrm{m}$
        \label{fig:pre_profile_shallow}
    \end{minipage}
    \hfill
    \begin{minipage}[t]{0.32\textwidth}
        \centering\small
        \begin{tabular}{cc}
            \textbf{Real XSEM} & \textbf{\shortstack{Calibrated\\Simulation}} \\
            \includegraphics[width=0.48\linewidth,height=4.0cm,keepaspectratio]{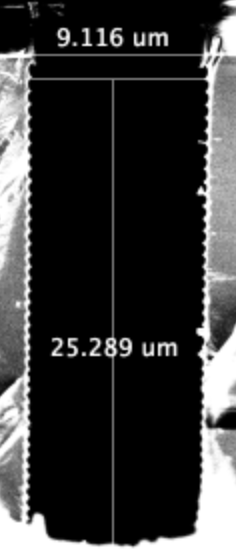} &
            \includegraphics[width=0.48\linewidth,height=4.0cm,keepaspectratio]{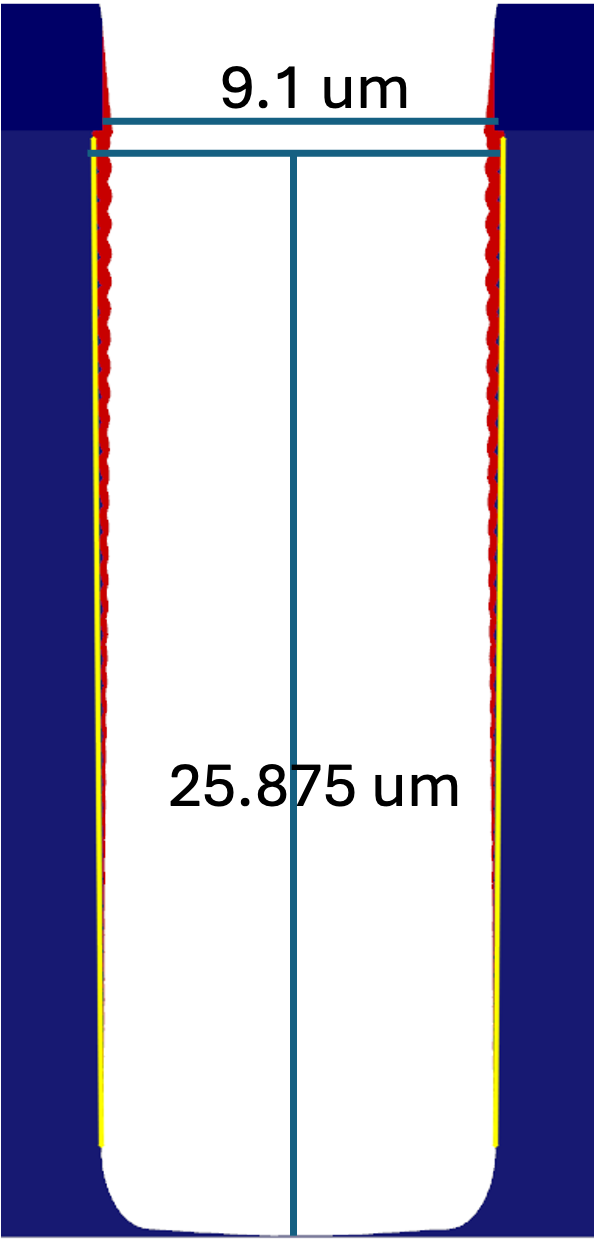}
        \end{tabular}
        \par\smallskip\textbf{(b)} Intermediate etching:
$c_a^{test}=40$, $o_a^{test}=9.116\,\mu\mathrm{m}$
        \label{fig:pre_profile_intermediate}
    \end{minipage}
    \hfill
    \begin{minipage}[t]{0.32\textwidth}
        \centering\small
        \begin{tabular}{cc}
            \textbf{Real XSEM} & \textbf{\shortstack{Calibrated\\Simulation}} \\
            \includegraphics[width=0.48\linewidth,height=4.0cm,keepaspectratio]{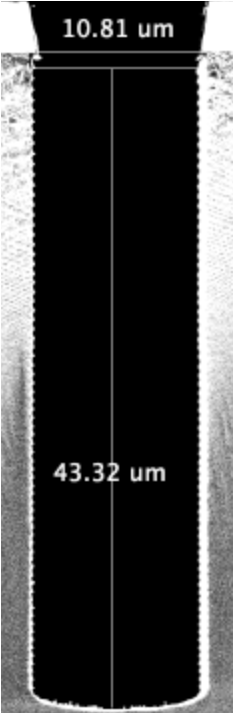} &
            \includegraphics[width=0.48\linewidth,height=4.0cm,keepaspectratio]{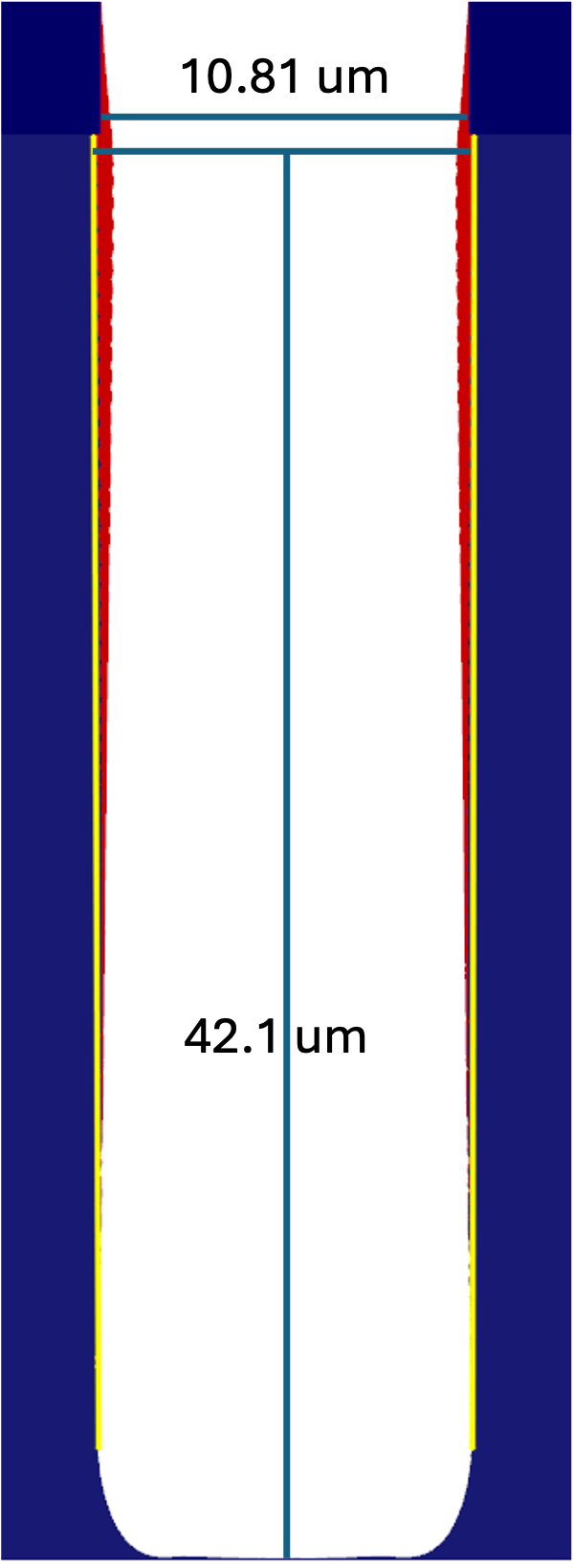}
        \end{tabular}
        \par\smallskip\textbf{(c)} Deep etching:
$c_a^{test}=70$, $o_a^{test}=10.81\,\mu\mathrm{m}$
        \label{fig:pre_profile_deep}
    \end{minipage}

    \caption{ Three representative prediction results on \(D^{(1)}_{\mathrm{test}}\) for Recipe 1. Covering shallow, intermediate, and deep etching. Each paired comparison shows the real XSEM profile on the left and the corresponding calibrated ViennaPS prediction on the right. The examples correspond to different cycle numbers and opening sizes in \(D^{(1)}_{\mathrm{test}}\), showing that the same calibrated parameter vector \(z_1^*\) can approximate the main real-profile geometry on test samples.}
    \label{fig:prediction_profiles}
\end{figure*}

Figure~\ref{fig:prediction_profiles} shows three representative prediction results from Recipe 1. Each paired comparison presents the real XSEM profile and the corresponding calibrated ViennaPS prediction generated using the recipe-specific parameter vector \(z_1^*\). For Recipe 1 examples, the calibrated simulator reproduces the main trench geometry, including the etch depth and sidewall direction, without sample specific parameter tuning. 
This result supports the quantitative results in Table~\ref{tab:prediction_summary}, suggesting that the identified shared parameter vector captures the dominant etching behavior beyond the fitting samples.

\begin{figure}[t]
\centering

\includegraphics[width=\linewidth]
{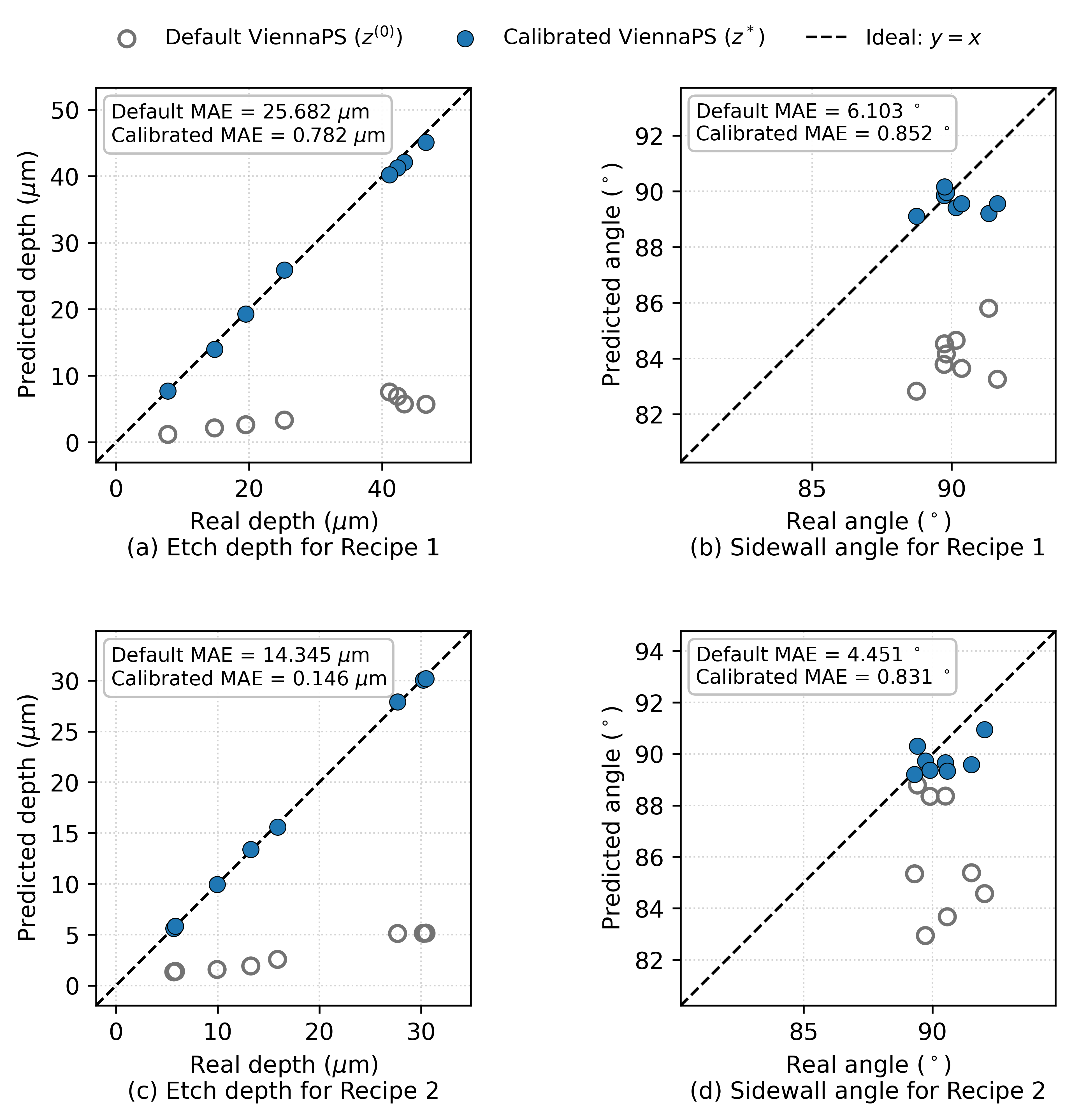}

\caption{
Real measurements versus predicted measurements on the held-out test sets of
Recipe 1 and Recipe 2.
The dashed line denotes the ideal \(y=x\) agreement.
Gray circles indicate the default ViennaPS results, and blue circles indicate
the calibrated ViennaPS results.
(a) Etch depth for Recipe 1.
(b) Sidewall angle for Recipe 1.
(c) Etch depth for Recipe 2.
(d) Sidewall angle for Recipe 2.
}
\label{fig:prediction_scatter}
\end{figure}

To examine prediction behavior across the full test set, Figure~\ref{fig:prediction_scatter} compares the predicted and real measurements on \(D^{(1)}_{\mathrm{test}}\) and \(D^{(2)}_{\mathrm{test}}\). For both recipes, the predicted etch depths remain close to the ideal \(y=x\) line, and the sidewall angle predictions also stay within a small error range around the real measurements. These results indicate that the calibrated simulator provides reliable ViennaPS predictions for held-out process conditions within their corresponding recipes.

\begin{figure}[t]
    \centering
    \includegraphics[width=1.0\linewidth]{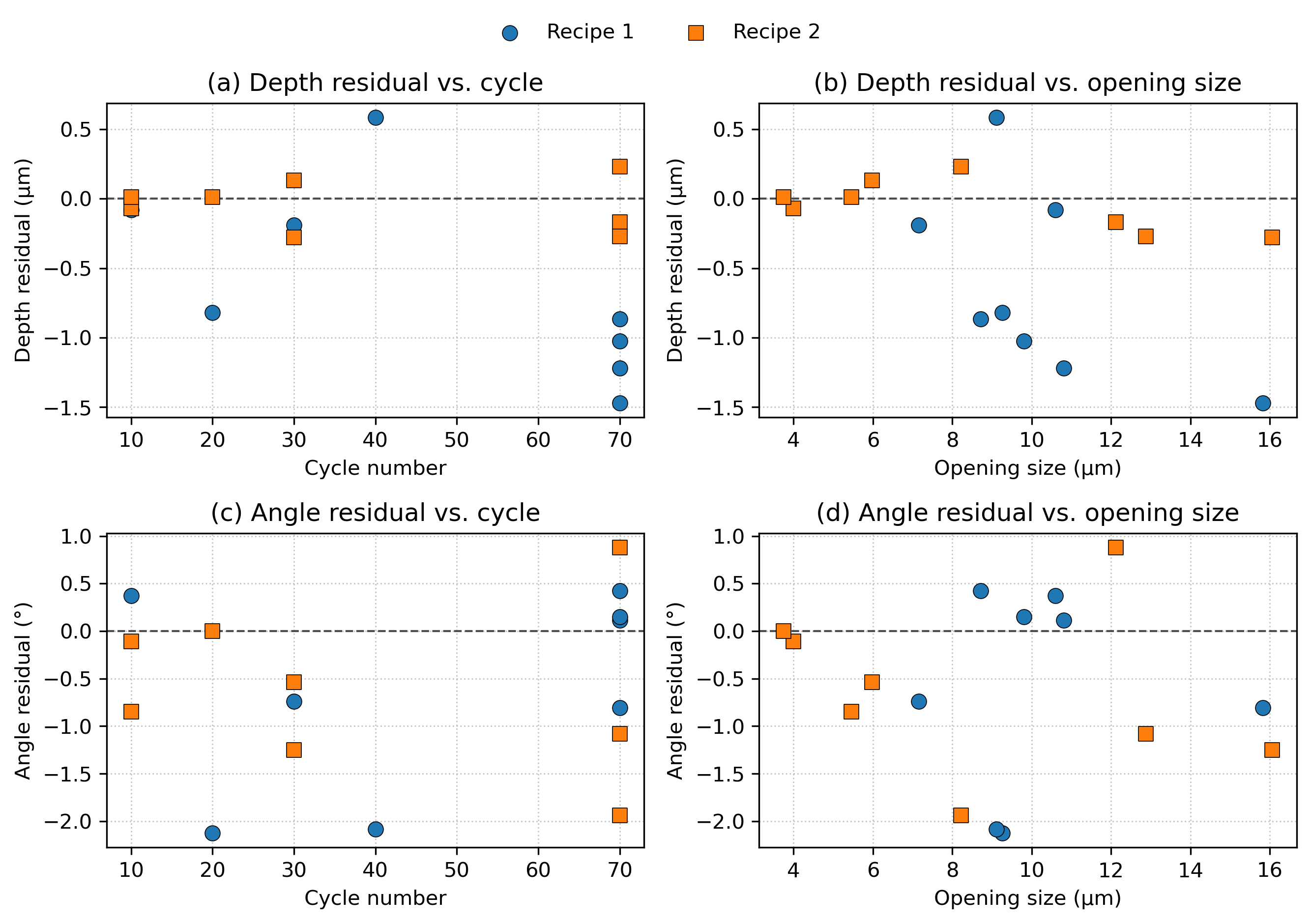}
    \caption{Residual analysis of prediction results on held-out test set for Recipe 1 (blue circles) and Recipe 2 (orange squares). Each marker denotes one test sample (\(N=8\) per recipe), and the residual is computed as the value predicted with the calibrated parameters \(z_r^*\) minus the real XSEM measurement. The residual is computed as predicted value minus real measurement. The dashed horizontal line denotes zero residual.}
    \label{fig:placeholder}
\end{figure}

To further examine whether the remaining prediction errors are systematic, Figure~\ref{fig:placeholder} shows residuals of etch depth and sidewall angle against cycle number and opening size. The residuals remain within a small range, suggesting that the calibrated parameter vectors $z_1^*$ and $z_2^*$ capture the dominant simulator-parameter discrepancy. 
The remaining local deviations likely reflect a combination of measurement noise in the real XSEM data and model form limitations inherent to the simulator.

Together with the quantitative metrics, representative profile comparisons, scatter plots, and residual analysis, these results show that the calibrated simulator provides consistent prediction performance on the corresponding held-out test set \(D^{(r)}_{\mathrm{test}}\).

\subsection{Ablation Study }
\label{sec:ablation}

This section shows how different candidate generation strategies affect calibration accuracy. To isolate the contribution of each component while controlling the experimental cost, the ablation study is conducted on Recipe 1 only. Since evaluating each candidate requires running ViennaPS on multiple fitting samples, the simulation cost increases with the number of search iterations. Therefore, this ablation study focuses on the low-cost calibration setting, where different candidate generation strategies are compared under the same 10-iteration budget. To examine whether the conclusions depend on this short search horizon, we additionally provide an extended 20-iteration comparison in Appendix ~\ref{app:20_iteration}

For a fair comparison, all variants use the same initial parameter vector $z_1^{(0)}$, parameter bounds, fitting set $D^{(1)}_{\mathrm{fit}}$, random seed, and number of evaluations per iteration. We compare three candidate generation strategies: local random exploration(Random only), local exploration with LLM guidance(Random+LLM), and local exploration with XGBoost screening(Random+XGBoost). Here, we adopt Qwen3-27B as the LLM backend. These are contrasted with the full proposed method(Ours), which combines all three components.

All candidates in a batch are generated and evaluated together before the next iteration begins.In other words, at iteration $t$, all candidate parameter vectors are generated from the search history available at the end of iteration $t-1$. The candidate composition of each variant is summarized in Table~\ref{tab:ablation_candidate_setting}. For the full method(Ours), each batch consists of three LLM-guided candidates, one local random candidate, and two surrogate-screened candidates. If the surrogate branch cannot provide enough valid candidates, the remaining slots are filled by local random exploration. The other variants use the same batch size but generate candidates differently, as shown in Table~\ref{tab:ablation_candidate_setting}. Once all six candidates are evaluated under the progressive evaluation scheme, the surrogate model and the current best parameter vector are updated at the end of the iteration.

\begin{table}[t]
\centering
\caption{Candidate composition for each ablation variant.}
\label{tab:ablation_candidate_setting}
\begin{tabular}{lcc}
\toprule
Variant & Candidate composition per iteration & Batch size \(B\)\\
\midrule
Random only & Random \(\times 6\) & 6 \\
Random + LLM & LLM \(\times 3\) + Random \(\times 3\) & 6 \\
Random + XGBoost & Random \(\times 3\) + XGBoost \(\times 3\) & 6 \\
Ours & LLM \(\times 3\) + Random \(\times 1\) + XGBoost \(\times 2\) & 6 \\
\bottomrule
\end{tabular}
\end{table}

To reduce the cost of exploration, we use a progressive evaluation strategy rather than evaluating every candidate on the full fitting set from the start. In the first stage, each candidate is evaluated on 10-cycle fitting samples. If its depth error on these samples does not exceed 10\%, it is promoted to the next stage and evaluated on two 20-cycle samples. Candidates that also satisfy the 20-cycle criterion are further evaluated on three 30-cycle samples. Starting from the 8th search iteration, all methods are evaluated on the complete fitting set $D^{(1)}_{\mathrm{fit}}$, regardless of prior promotion status. Such a design reduces the number of evaluations required in the early stages by screening candidates on shorter cycle samples first. The first seven iterations are used for low-cost exploration, while the final three use the full fitting set for verification and refinement. Poor candidates are rejected early using inexpensive low-cycle simulations, and the concluding full-set evaluation ensures that the selected parameter vector is validated across all fitting regimes. Therefore, the reported search cost is measured by the accumulated number of ViennaPS sample-level evaluations.

\begin{figure}[t]
    \centering
    \includegraphics[width=0.7\linewidth]{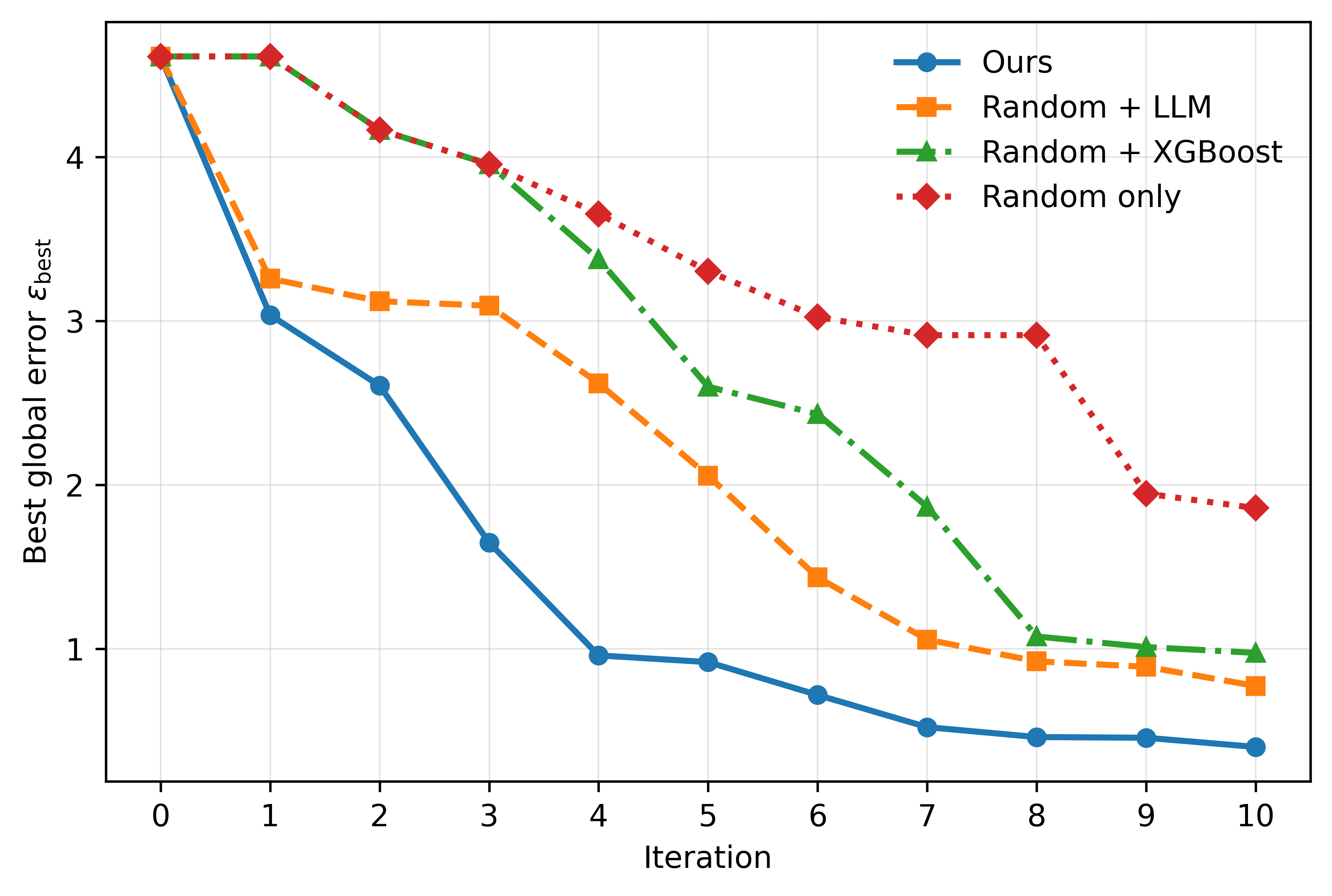}
    \caption{Ablation study on candidate generation strategies using Recipe 1, with qwen3.8 used as the LLM for candidate generation.}
    \label{fig:ablation}
\end{figure}

Figure~\ref{fig:ablation} shows the convergence of different candidate generation strategies. The x-axis denotes the iteration number and the y-axis denotes the best global error \(\epsilon_{\mathrm{best}}\), which is defined in Eq.~\ref{global_error}. All methods are compared using the same complete fitting set, even though their intermediate candidate screening may use progressive subsets of \(D^{(1)}_{\mathrm{fit}}\). The plotted error reflects the best validated calibration performance achieved up to each iteration, and a faster decrease indicates that strategy identifies a low error parameter region with fewer iterations. 

As shown in Figure~\ref{fig:ablation}, our method achieves the fastest reduction of \(\epsilon_{\mathrm{best}}\), which indicates that combining LLM-guided proposals, local random exploration, and surrogate screening helps the search enter a promising parameter region early and continue refining it.
Random+LLM guidance variant also converges faster than Random only, showing that LLM guidance provides useful directional updates beyond purely random perturbations. 
Random+XGBoost variant decreases slowly at the beginning, but drops more clearly after several iterations. This is likely because the surrogate model requires sufficient evaluated candidates before it can reliably rank new parameter vectors. In the first three iterations, this variant has the same error curve as the Random only variant due to the surrogate warm-up setting. Specifically, XGBoost screening is activated only after the search history contains at least 36 sample-level evaluated data points. Since each of the first three iterations evaluates a batch of six candidates on two 10-cycle fitting samples, the search accumulates \(3 \times 6 \times 2 = 36\) sample evaluations by the end of the third iteration. Before this threshold is reached, the surrogate is filled by local random exploration. Therefore, during the warm-up phase, Random+XGBoost screening variant is equivalent to the local random exploration variant. To avoid redundant ViennaPS runs under identical candidate generation behavior, the same warm-up trajectory is used for both variants. 
Random only variant decreases slowly and remains at a higher error, showing that random local exploration alone is inefficient in this calibration problem.

\begin{table}[t]
\centering
\caption{Ablation study on candidate generation strategies.}
\label{tab:ablation_efficiency}
\begin{tabular}{lcccc}
\toprule
Method & Local & LLM & XGBoost & $\epsilon_{\mathrm{final}}(D^{(1)}_\mathrm{fit})$ \\
\midrule
Random only      & \checkmark & --         & --         & 1.858 \\
Random + LLM     & \checkmark & \checkmark & --         & 0.772 \\
Random + XGBoost & \checkmark & --         & \checkmark & 0.974 \\
Ours            & \checkmark & \checkmark & \checkmark & 0.400 \\
\bottomrule
\end{tabular}
\end{table}

Table~\ref{tab:ablation_efficiency} summarizes the final calibration error of different candidate generation strategies. Local random exploration variant gets the highest final error, indicating that purely random perturbation is insufficient for this calibration task. Adding LLM guidance reduces the final error from 1.858 to 0.772, showing that LLM proposals provide useful search directions beyond random exploration. Adding XGBoost screening reduces the final error to 0.974, suggesting that the surrogate model helps filter candidates once sufficient evaluation history is available. The full method achieves the lowest final error of 0.400, outperforming all variants. These results suggest that LLM guidance and XGBoost screening play complementary roles. The LLM steers the search toward promising regions, while the surrogate improves candidate selection based on accumulated evaluations.

Overall, the ablation results show that the three candidate generation components play complementary roles in different stages of calibration. Local exploration provides basic parameter space coverage, but it is inefficient when used alone. LLM improves early search, while XGBoost screening becomes useful after enough simulation data has been accumulated. By combining these components, the full method achieves both fast early error reduction and the lowest final calibration error under the same number of iterations. These results confirm that the performance gain comes from the hybrid search design rather than from any single candidate generation source.

\subsection{LLM Robustness Analysis}

The Section \ref{sec:ablation} shows that incorporating the LLM-guided candidate generation module significantly improves the calibration performance. In this section, we investigate whether such improvement depends on the choice of a specific LLM. To evaluate the robustness of the proposed framework across different LLM backends, we keep the other modules, search prompts, and calibration data unchanged. In this experiment, we use the data from Recipe 1.

For each LLM, we run the complete parameter calibration process separately to obtain the corresponding optimal simulator parameter vector$z^*$. Then, we apply the different $z^*$ to the same set of held out test process conditions. Finally, we adopt the global normalized error, the MAE of etch depth and the MAE of sidewall angle as evaluation metrics.

\begin{table}[t]
\centering
\caption{Robustness analysis across different LLM backends.}
\label{tab:llm_robustness}
\begin{tabular}{lccc}
\toprule
LLM Backend
& \(\mathrm{MAE}_d\)  ($\mu$m)
& \(\mathrm{MAE}_a\) ($^\circ$)
& $\epsilon(z^*;D_{\mathrm{test}})$\\
\midrule
gpt-oss:120b   & 0.681 & 0.915 & 0.344 \\
qwen3.8:27b    & 0.782 & 0.852 & 0.340 \\
gemma4:26b-a4b & 0.754 & 1.058 & 0.401 \\
llama4:latest  & 0.302 & 1.144 & 0.334 \\
\bottomrule
\end{tabular}
\end{table}

Table \ref{tab:llm_robustness} shows the robustness under different LLM backends. The results show that gpt-oss:120b, qwen3.8:27b, gemma4:26b-a4b and llama4:latest achieve very similar errprs on the held-out test set. Although the $z^*$ predicted by different LLMs shows some differences in two metrics. llama4:latest achieves the lowest depth MAE(0.302$\mu$m),but its sidewall angle MAE is relatively high(1.144$^\circ$). In contrast, qwen3.8:27b has a depth MAE of (0.782$\mu$m), but achieves the lowest angle MAE(0.852$^\circ$). This shows that different LLMs may explore different error trade-offs between depth and sidewall angle, while the final overall calibration error stays similar. Therefore, the overall calibration performance of the framework does not strongly depend on a specific LLM backend.

\subsection{Fitting Data Size Sensitivity}

To examine how the number of real calibration samples affects the identified simulator parameter, we vary the fitting set size as $N_{\mathrm{fit}}\in\{6,8,10,12,14\}$. This sensitivity analysis is conducted on Recipe 1 only and each split uses 6 held-out test samples, leaving 14 samples available for calibration. The experiment is repeated over three independent data splits. In each split, a test set is first held out and kept unchanged for all \(N_{\mathrm{fit}}\) settings within that split. The fitting subsets of different sizes are then selected from the remaining samples. For each \(N_{\mathrm{fit}}\), the calibration framework is run independently on each split, producing a separate calibrated parameter vector. The mean and standard deviation of the test depth MAE, sidewall angle MAE, and normalized global error are reported over the three splits.

\begin{figure}[t]
    \centering
    \includegraphics[width=1.0\linewidth]{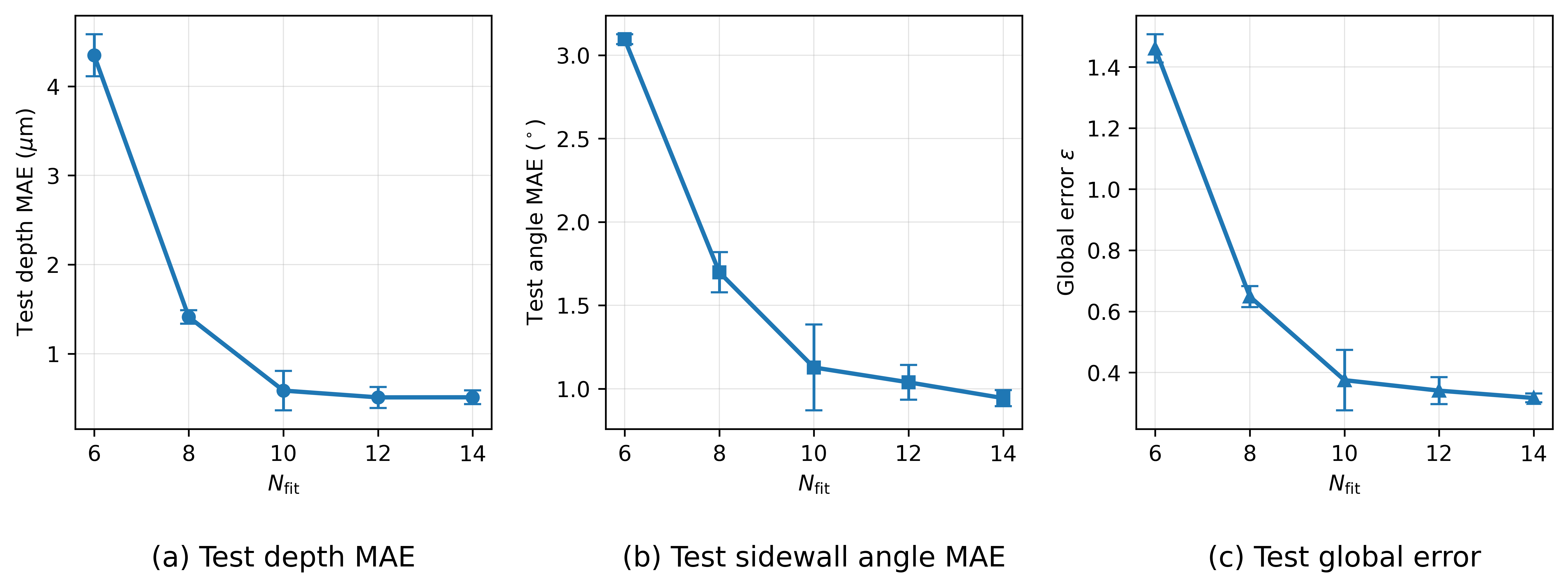}
    \caption{Fitting data size sensitivity of the calibration framework. For each \(N_{\mathrm{fit}}\), the complete calibration framework is independently run over three data splits. Within each split, the same held-out test set is used for all \(N_{\mathrm{fit}}\) settings. Markers denote the mean test performance over the three runs, and error bars indicate one standard deviation across runs.}
    \label{fig:data_size_sensitivity}
\end{figure}

Figure~\ref{fig:data_size_sensitivity} shows the fitting data-size sensitivity of our calibration framework. The x-axis is the number of real XSEM samples used for calibration, \(N_{\mathrm{fit}}\), and the y-axis is the depth MAE, angle MAE and prediction global error on the held-out test set. For each \(N_{\mathrm{fit}}\), we run the complete calibration framework three times using independent data splits. 
In each run, the test set is kept fixed across all \(N_{\mathrm{fit}}\) settings.
The markers denote the mean test performance over the three runs, and the error bars indicate one standard deviation across runs. A lower test error indicates that the calibrated simulator predicts held-out process conditions within the same recipe more accurately. As \(N_{\mathrm{fit}}\) increases, the test depth MAE, test sidewall angle MAE, and test global error all decrease. Beyond \(N_{\mathrm{fit}}=10\), the curves flatten and the gains become marginal, indicating that the calibrated simulator has already reached a relatively stable regime. 

Table~\ref{tab:data_size_sensitivity} reports the fitting data size sensitivity averaged over three independent runs. With only six samples, the calibrated simulator still has large prediction errors on the test samples. As $N_{\mathrm{fit}}$ increases from 6 to 10, all three metrics decrease substantially, showing that adding a small number of real XSEM samples improves the identified simulator parameter. For $N_{\mathrm{fit}} \geq 10$, the improvement becomes more gradual. This indicates that the calibration enters a relatively stable regime once about 10 fitting samples are available, and further increasing the fitting set provides only marginal additional gains. Overall, these results confirm that stable calibration is achievable with a small number of real XSEM measurements.

\begin{table}[t]
\centering
\caption{Fitting data size sensitivity averaged over three independent runs.}
\label{tab:data_size_sensitivity}
\begin{tabular}{cccc}
\toprule
$N_{\mathrm{fit}}$ & Test $\mathrm{MAE}_d$ ($\mu\mathrm{m}$) & Test $\mathrm{MAE}_{\alpha}$ ($^\circ$) & $\epsilon(z_N^*;D_{\mathrm{test}})$ \tabularnewline
\midrule
6  & 4.348 & 3.096 & 1.461 \\
8  & 1.412 & 1.698 & 0.649 \\
10 & 0.586 & 1.127 & 0.375 \\
12 & 0.510 & 1.038 & 0.341 \\
14 & 0.511 & 0.943 & 0.317 \\
\bottomrule
\end{tabular}
\end{table}

\subsection{Batch Size Sensitivity}

This section investigates how the number of candidate simulator parameter vectors evaluated per iteration, referred to as the batch size, affects the identified simulator parameter. This sensitivity analysis is conducted on Recipe 1 only. To ensure a fair comparison, all batch size settings are compared under the same budget of 60 candidate evaluations. Each candidate evaluation corresponds to testing one simulator parameter vector. We evaluate batch sizes of 4, 5, 6, 10, and 12, corresponding to 15, 12, 10, 6, and 5 search iterations. All settings use the same initial parameter vector, fitting set, parameter bounds, and candidate generation strategy. The best global error is reported as the lowest fitting error found within the 60 candidate evaluation budget.

\begin{figure}
    \centering
    \includegraphics[width=0.75\linewidth]{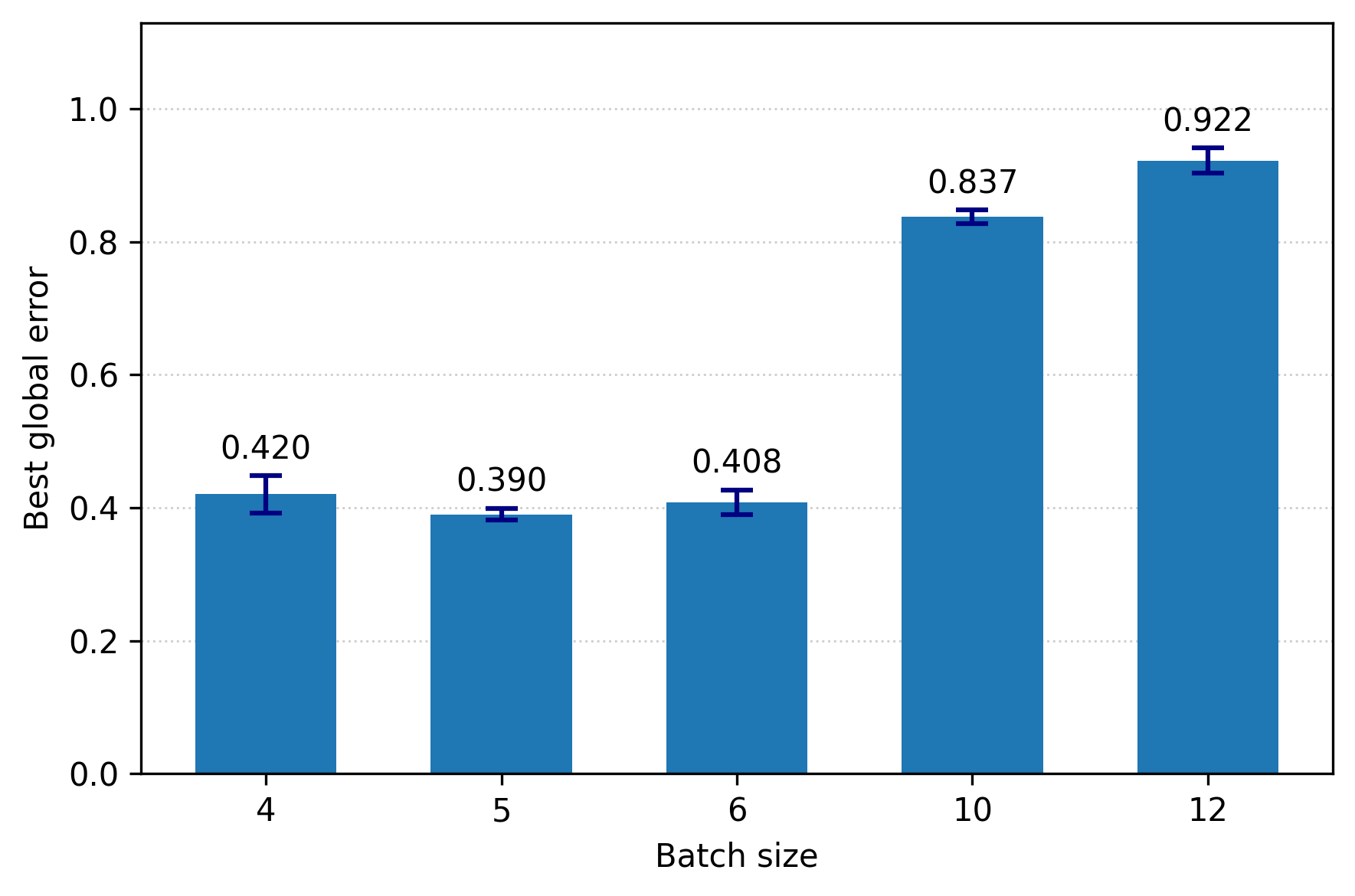}
    \caption{Batch size sensitivity under a fixed simulation budget of $60$ candidate evaluations. Bars denote the mean final best global error over three independent runs, and error bars indicate one standard deviation.}
    \label{fig:batch_size_sensitivity}
\end{figure}

Figure~\ref{fig:batch_size_sensitivity} shows the effect of batch size under a fixed budget of $60$ candidate evaluations. Each bar shows the mean final best global error over three independent runs and the error bar indicates one standard deviation. The best global errors for batch size equal to $4$, $5$, and $6$ are $0.420$, $0.390$, and $0.408$, respectively, indicating that the calibration method is insensitive to the batch size within a moderate range. The slight decrease from batch size 4 to 5 indicates that an overly small batch may limit candidate diversity within each iteration. When batch size is 4, the number of candidates evaluated before etch update is reduced, which may limit the likelihood of identifying a more favorable parameter direction. Increasing the batch size to 5 slightly improves candidate diversity while still has frequent updates. However, when the batch size increases to $10$ and $12$, the best global errors increase to 0.837 and 0.922. This result is likely because larger batches reduce the number of feedback updates under the same evaluation budget. For example, under the same budget of 60 candidate evaluations, batch size 6  allows 10 feedback updates, while batch size 10 allows only 6 updates. This affects both LLM-guided proposal and XGBoost-based screening. When the current search direction is not ideal, a large batch may spend many candidate evaluations before the LLM can correct its next proposal, and many candidates are screened against an older surrogate model before the newly evaluated simulation records can be incorporated. Overall, a moderate batch size is preferred because it maintains sufficient feedback updates while still providing enough candidate diversity in each iteration.

\section{Conclusion and Future Work}\label{sec:conclusion}
In this work, we propose a few-shot real-to-sim calibration framework for ViennaPS targeting the Bosch process. The key idea is to use a small number of real XSEM measurements to identify a recipe-specific simulator parameter vector \(z_r^*\) that aligns simulated etch depth and sidewall angle with real Bosch etching results across different cycle numbers and opening sizes within each recipe. The method combines local exploration, LLM-guided candidate generation, and XGBoost-based surrogate screening, while keeping ViennaPS simulation as the final verification step. This design is particularly suitable for Bosch etching calibration with sparse terminal XSEM measurements. And it requires neither large-scale real labeled data, nor fully specified Bosch plasma-chemistry equations, nor continuous observations such as trajectories or image sequences. Experiments show that the calibrated \(z_r^*\) substantially reduces fitting error and generalizes to test samples, indicating that the proposed framework can be consistently applied across different Bosch recipe settings using only a small number of real XSEM measurements.

Future work will focus on high-fidelity simulator calibration. Our current framework operates within a relatively low-dimensional parameter space. However, future high-fidelity simulators may introduce substantially more parameters, rendering calibration a challenging high-dimensional optimization problem. Therefore, we plan to investigate scalable methods that remain effective even when real XSEM data are limited.

\backmatter

\begin{appendices}

\section{Real Experiment Recipe and Samples} \label{app:recipe_sample}
Table~\ref{tab:real_recipe} shows the two Bosch process recipes used to fabricate the real etching samples in this work. Each recipe defines a fixed set of process parameters, while the cycle number, GDS design, and opening size vary across samples within each recipe. Table~\ref{tab:full_real_samples} provides the complete list of 20 real Bosch etching samples for Recipe 1, including the cycle number, opening size, measured etch depth, measured sidewall angle, and the corresponding data split. Among these, 12 samples are assigned to $D_{\text{fit}}$ for parameter identification, and the remaining 8 are held out as $D_{\text{test}}$ for final evaluation.

\begin{table}[t]
\centering
\caption{Bosch process recipes used for the real etching samples.}
\label{tab:real_recipe}

\small
\setlength{\tabcolsep}{4pt}

\begin{tabular}{lccccccc}
\toprule
Step & C$_4$F$_8$ (sccm) & SF$_6$ (sccm) & Ar (sccm) &
Pressure (mTorr) & Time (s) & ICP (W) & Bias (W) \\
\midrule

\multicolumn{8}{c}{\textbf{Recipe 1}} \\
\midrule
Ignition   & --  & 75  & 30 & 20 & 10 & 1500 & 100 \\
Deposition & 150 & --  & 30 & 25 & 2  & 2000 & 5   \\
Etch A     & --  & 150 & 30 & 60 & 2  & 2000 & 50  \\
Etch B     & --  & 150 & 30 & 60 & 2  & 2500 & 5   \\

\midrule
\multicolumn{8}{c}{\textbf{Recipe 2}} \\
\midrule
Ignition   & -- & 75 & 30 & 30 & 10 & 1500 & 100 \\
Deposition & 150 & -- & 30 & 30 & 2 & 2000 & 5  \\
Etch A     & -- & 150 & 30 & 35 & 2 & 2000 & 50 \\
Etch B     & -- & 100 & 30 & 35 & 2 & 2000 & 5 \\

\bottomrule
\end{tabular}
\end{table}

\begin{table}[h!]
\centering
\small
\caption{Full list of real Bosch etching samples for Recipe 1.}
\label{tab:full_real_samples}
\begin{tabular}{ccccc}
\toprule
Cycle number & Opening size ($\mu$m) & Real depth ($\mu$m) & Real angle ($^\circ$) & Split \\
\midrule
10 & 6.200 & 7.050 & 88.82 & $D_{\text{fit}}$ \\
10 & 10.600 & 7.780 & 88.74 & $D_{\text{test}}$  \\
10 & 11.000 & 7.840 & 90.91 & $D_{\text{fit}}$ \\
20 & 7.589 & 14.152 & 91.49 & $D_{\text{fit}}$  \\
20 & 8.344 & 14.770 & 89.73 & $D_{\text{fit}}$ \\
20 & 9.263 & 14.790 & 91.33 & $D_{\text{test}}$ \\
30 & 7.150 & 19.490 & 90.16 & $D_{\text{fit}}$ \\
30 & 8.120 & 19.930 & 89.94 & $D_{\text{fit}}$ \\
30 & 9.100 & 20.310 & 90.31 & $D_{\text{test}}$ \\
30 & 9.858 & 20.475 & 90.15 & $D_{\text{fit}}$ \\
40 & 7.552 & 23.950 & 91.33 & $D_{\text{fit}}$\\
40 & 8.256 & 24.820 & 90.54 & $D_{\text{fit}}$ \\
40 & 9.116 & 25.289 & 91.64 & $D_{\text{test}}$ \\
40 & 10.193 & 25.732 & 91.12 & $D_{\text{fit}}$ \\
70 & 6.950 & 38.860 & 89.60 & $D_{\text{fit}}$ \\
70 & 7.640 & 40.090 & 90.00 & $D_{\text{test}}$  \\
70 & 8.720 & 41.090 & 89.74 & $D_{\text{test}}$ \\
70 & 9.800 & 42.300 & 89.81 & $D_{\text{test}}$ \\
70 & 10.810 & 43.320 & 89.73 & $D_{\text{test}}$ \\
70 & 15.820 & 46.570 & 90.36 & $D_{\text{fit}}$ \\
\bottomrule
\end{tabular}
\end{table}

\section{ViennaPS Recipe}
\label{app:Vienna_recipe}

This section summarizes the ViennaPS Bosch ray-tracing configuration used in this work. For each real sample, the cycle number \(c_i\) and opening size \(o_i\) are set according to the corresponding measured process condition. 
These two quantities are treated as sample specific process inputs rather than calibration variables. During calibration, the optimizer only updates the eight flux-related parameters in the simulator parameter vector \(z\). All remaining ViennaPS settings are kept fixed for all candidate parameter vectors.
The geometry, numerical settings, process-step template, sticking probabilities, and effective reaction coefficients are kept fixed across all simulations, as listed in Tables~\ref{tab:fixed_geometry_numerical}-~\ref{tab:fixed_reaction_coefficients}. The eight flux-related parameters listed in Table~\ref{tab:initial_flux_parameters} are updated during calibration.

\begin{table}[!htbp] 
\centering 
\caption{Fixed geometry and numerical settings used in ViennaPS.} \label{tab:fixed_geometry_numerical} 
\begin{tabular}{lll} 
\hline Parameter & Value & Unit / note \\ 
\hline 
Grid spacing, \texttt{gridDelta} & 0.025 & \(\mu\mathrm{m}\) \\  
Mask height, \texttt{maskHeight} & 5.6 & \(\mu\mathrm{m}\) \\ 
Taper angle, \texttt{taperAngle} & 0.0 & degree \\ 
Ion source exponent, \texttt{ionSourceExponent} & 6000 & -- \\ 
Etch-step sequence, \texttt{etchStep} & both & Etch A followed by Etch B \\ 
\hline 
\end{tabular} 
\end{table}

\begin{table}[!htbp] 
\centering 
\caption{Fixed process template used in the ViennaPS Bosch simulation. The ignition stage is applied only once at the beginning of the simulation for initial trench opening, whereas the deposition, Etch A, and Etch B steps are repeated according to the sample specific cycle number.} 
\label{tab:fixed_process_template} 
\begin{tabular}{llllll}
\hline
Step & Main gas & Pressure (R1) & Pressure (R2) & Time & Role \\
\hline
Ignition & SF$_6$ / Ar & 20 mTorr & 30 mTorr & 10 s & Initial trench opening \\
Deposition & C$_4$F$_8$ / Ar & 25 mTorr & 30 mTorr & 2 s & Sidewall passivation \\
Etch A & SF$_6$ / Ar & 60 mTorr & 35 mTorr & 2 s & Main etching step \\
Etch B & SF$_6$ / Ar & 60 mTorr & 35 mTorr & 2 s & Secondary etching step \\
\hline
\end{tabular}
\end{table}

\begin{table}[!htbp] 
\centering 
\caption{Fixed sticking probabilities used in the ViennaPS Bosch simulation.} 
\label{tab:fixed_sticking} 
\begin{tabular}{lll} \hline Step & Parameter & Value \\ 
\hline 
Ignition & \texttt{stickingIgnitionMask} & 0.2 \\ 
Ignition & \texttt{stickingIgnitionSubs} & 0.2 \\ 
Deposition & \texttt{stickingDep} & 0.1 \\ 
Etching & \texttt{stickingEtchPoly} & 0.1 \\ 
Etching & \texttt{stickingEtchMask} & 0.2 \\ 
Etching & \texttt{stickingEtchSubs} & 0.2 \\ 
\hline 
\end{tabular} 
\end{table} 

\begin{table}[!htbp] 
\centering 
\caption{Fixed effective reaction coefficients used in the ViennaPS Bosch simulation.} 
\label{tab:fixed_reaction_coefficients} 
\begin{tabular}{llll} 
\hline 
Step & Parameter & Value & Physical meaning \\ 
\hline 
Ignition & \texttt{alphaIgnitionSubs} & -270 & Ion-assisted etching of Si \\ 
Ignition & \texttt{alphaIgnitionMask} & -13.5 & Ion-assisted etching of mask \\ 
Ignition & \texttt{betaIgnitionSubs} & -0.9 & Chemical etching of Si \\ 
Ignition & \texttt{betaIgnitionMask} & -0.045 & Chemical etching of mask \\ 
Deposition & \texttt{alphaDep} & 10.0 & Ion-assisted deposition coefficient \\ 
Deposition & \texttt{betaDep} & 0.5 & Neutral-assisted deposition coefficient \\ 
Etching & \texttt{alphaPoly} & -125 & Ion-assisted etching of polymer \\ 
Etching & \texttt{alphaSubs} & -270 & Ion-assisted etching of Si \\ 
Etching & \texttt{alphaMask} & -13.5 & Ion-assisted etching of mask \\ 
Etching & \texttt{betaPoly} & -0.03 & Chemical etching of polymer \\ 
Etching & \texttt{betaSubs} & -0.9 & Chemical etching of Si \\ 
Etching & \texttt{betaMask} & -0.045 & Chemical etching of mask \\ 
\hline 
\end{tabular} 
\end{table}

\begin{table}[!htbp] 
\centering 
\caption{Flux-related simulator parameters updated during calibration. The listed values correspond to the initial ViennaPS configuration before calibration.} 
\label{tab:initial_flux_parameters} 
\begin{tabular}{llll} 
\hline 
Parameter & Step & Initial value & Unit \\ 
\hline
\(\mathrm{Flux}_{ion}^{I}\) & Ignition & \(2.50\times10^{15}\) & atoms/cm\(^2\cdot\)s \\ 
\(\mathrm{Flux}_{neu}^{I}\) & Ignition & \(6.00\times10^{18}\) & atoms/cm\(^2\cdot\)s \\
\(\mathrm{Flux}_{ion}^{D}\) & Deposition & \(1.40\times10^{15}\) & atoms/cm\(^2\cdot\)s \\ 
\(\mathrm{Flux}_{neu}^{D}\) & Deposition & \(3.50\times10^{18}\) & atoms/cm\(^2\cdot\)s \\ 
\(\mathrm{Flux}_{ion}^{EA}\) & Etch A & \(4.00\times10^{16}\) & atoms/cm\(^2\cdot\)s \\ 
\(\mathrm{Flux}_{neu}^{EA}\) & Etch A & \(1.80\times10^{19}\) & atoms/cm\(^2\cdot\)s \\ 
\(\mathrm{Flux}_{ion}^{EB}\) & Etch B & \(2.54\times10^{16}\) & atoms/cm\(^2\cdot\)s \\ 
\(\mathrm{Flux}_{neu}^{EB}\) & Etch B & \(1.50\times10^{19}\) & atoms/cm\(^2\cdot\)s \\ 
\hline 
\end{tabular} 
\end{table}

\section{LLM-Based Parameter Guidance Prompt}
\label{app:llm_prompt}
The LLM is used as a candidate parameter guidance module. At each iteration, the system message defines the editable parameters, physical interpretation rules, and output constraints. The user message provides the current calibration state, including the current parameters value, simulated profile, real measurement, residuals, and simulated state. The prompt used for LLM-based candidate generation is provided below.

\begin{lstlisting}[basicstyle=\ttfamily\small,breaklines=true]
System message:

You are a Bosch deep silicon etching (SF6/C4F8 alternating)
ray-tracing simulation calibration assistant. Your goal is to
reduce the gap between simulation and measurement by adjusting
only nominal flux parameters Flux_*.

You may suggest new numeric values only for the following keys:
Flux_ionI, Flux_neuI, Flux_ionD, Flux_neuD, Flux_ionE_A, Flux_neuE_A, Flux_ionE_B, Flux_neuE_B.

Do not change alpha*, beta*, sticking*, ionSourceExponent, step times, geometry, numCycles, or etchStep.

Each proposal must include a short physical reason. The reason should explain why the proposed flux change may reduce the current depth or sidewall-angle error.

Inputs:
- z: current Flux_* values.
- S(z): simulated depth d_sim and sidewall angle alpha_sim.
- y: real measured depth d_real and sidewall angle alpha_real.
- r = y - S(z): residuals r_d and r_alpha.
- angle convention: definition of the sidewall angle.

Residual interpretation:
- If r_d > 0, the simulated depth is too shallow. Prefer increasing etching fluxes or reducing excessive protection.
- If r_d < 0, the simulated depth is too deep. Prefer decreasing etching fluxes or increasing deposition-related fluxes.
- The meaning of r_alpha depends on the angle convention. Use the provided angle definition to decide whether the simulated sidewall is too vertical or too laterally etched.

Physical guidance:
- Ignition fluxes Flux_ionI and Flux_neuI mainly affect the initial opening and early profile.
- Deposition fluxes Flux_ionD and Flux_neuD mainly affect polymer deposition and sidewall protection.
- Etching fluxes Flux_ionE_A/B and Flux_neuE_A/B control the etching stages. Ion fluxes mainly strengthen directional ion-assisted etching. Neutral fluxes mainly strengthen chemical or more isotropic etching.

Output requirements:
Return 1--3 candidate updates in strict JSON format. Each candidate should modify only a small number of flux parameters.
Each candidate must contain:
- changes: explicit numeric values for modified Flux_* keys.
- magnitude: small, medium, or large for each changed key.
- reason: a short physical explanation.

Keep all forbidden parameters unchanged.
User message at iteration t:

Angle convention:
<definition of sidewall angle>

Current flux vector z^(t):
{
  "Flux_ionI": ...,
  "Flux_neuI": ...,
  "Flux_ionD": ...,
  "Flux_neuD": ...,
  "Flux_ionE_A": ...,
  "Flux_neuE_A": ...,
  "Flux_ionE_B": ...,
  "Flux_neuE_B": ...
}

Simulation S(z): d_sim = ... , alpha_sim = ...
Real measurement y: d_real = ... , alpha_real = ...
Residual r = y - S(z): r_d = ... ,r_alpha = ...

Global calibration information: global_error = ... , validity_label = ... , worst_sample_residual = ...

Task:
Analyze the current depth error, sidewall-angle error, and
invalid simulation cases. Propose flux-only candidate updates
for the next iteration.

Return strict JSON only:
{
  "candidates": [
    {
      "name": "strategy_1",
      "changes": {
        "Flux_ionE_A": ...
      },
      "magnitude": {
        "Flux_ionE_A": "small"
      },
      "reason": "..."
    }
  ]
}

\end{lstlisting}

\section{Extended 20-Iteration Ablation Study}
\label{app:20_iteration}

Since each calibration iteration requires multiple ViennaPS evaluations, the main ablation study focuses on the low-cost 10-iteration setting. To examine whether the conclusions are sensitive to this short search horizon, we further extend all ablation variants to 20-iteration under the same experimental setting.

\begin{figure}[t]
\centering
\includegraphics[width=0.7\linewidth]{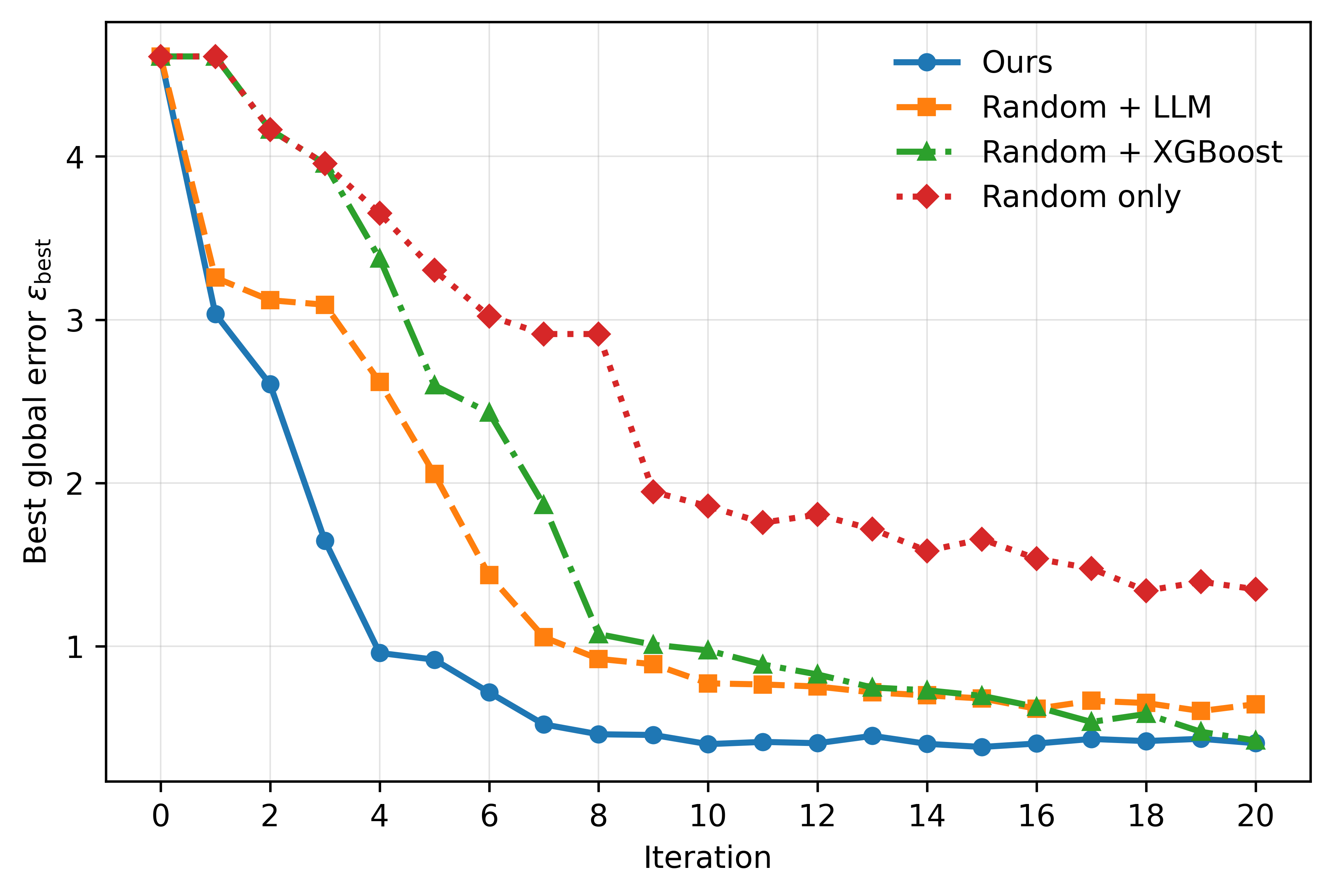}
\caption{Extended 20-iteration ablation study on candidate generation strategies. All variants follow the same experimental protocol as the main 10-iteration ablation study.}
\label{fig:extended_ablation_20iter}
\end{figure}

Figure~\ref{fig:extended_ablation_20iter} shows the extended 20-iteration ablation results. Random only remains the weakest variant, showing that random local exploration alone is inefficient for this calibration problem. With more iterations, Random+LLM and Random+XGBoost continue to improve, but these methods require a larger simulation budget to approach the error level that our method reaches earlier. Overall, these results support that the hybrid search achieves good search efficiency and can still produce a high quality calibration result.

\end{appendices}

\clearpage
\bibliography{sn-bibliography}

\end{document}